# Structured light interferometry

Liang Fang, Zhenyu Wan, Jian Wang[*]

Wuhan National Laboratory for Optoelectronics, School of Optical and Electronic Information, Huazhong University of Science and Technology, Wuhan 430074, Hubei, China.

* Correspondence to: jwang@hust.edu.cn

**Optical interferometry has made tremendous development in fundamentals and applications of light. Traditional optical interferometry using simplex plane-phase light just allows one-dimensional measurement, but incapable of retrieving multi-dimensiona1 movement. Here, we present new structured light interferometry to achieve simultaneous determination of rotational and translational dimensionalities (magnitudes and directions) of moving objects by fully exploiting multiple degree of freedoms of structured light. Such multi-dimensiona1 detectability benefits from multiple amplitude- and sign-distinguishable Doppler shifts of structured light by analyzing relative amplitude and phase information of Doppler signals. We experimentally demonstrate two interferometric schemes to individually retrieve both translational and rotational velocity vectors of two kinds of moving objects (small particles and bulk targets). These interferometric schemes based on structured light show remarkable function enhancement for simultaneous measurement of multi-dimensional motion vectors, compared with traditional interferometry, and thus may expand optical interferometry to multi-dimensional metrology in remote sensing, intelligent manufacturing and engineering.**

Optical interferometry, one of the most sensitive metrics tool, has gained rapid development over the past century. It is of great importance in advanced science and technology and allows numerous applications in optical metrology, nanophotonics, quantum physics, astronomy, remote sensing, industrial engineering and intelligent manufacturing, etc[1-9]. As a great example, the giant Michelson interferometer of LIGO was successfully built to achieve the detection of gravitational wave[10,11]. So far, optical interferometers were developed into myriad categories as diverse as Michelson interferometer, Mach-Zehnder interferometer, Twyman-Green interferometer, Sagnac interferometer, point diffraction interferometer, Rayleigh interferometer[12-17]. Even so, almost all optical interferometers just utilize the simplex phase degree of freedom (DoF) of light to determine simplex physical information, such as linear motion, rotating angle, planeness, straightness, or perpendicularity.

In a standard interferometric scheme, the Gaussian beam employed only allows detection of the translational (longitudinal) movement component of moving objects along the axis of light, but incapable of determining its perpendicular (transverse) component. In order to detect this dimensionality, an applaudable approach is to use the twisted or vortex beam[18-20], because its azimuthal phase gradient enables interaction with the transverse (or rotational) velocity. The underlying mechanism of translation- and rotation-detectable measurement is associated with the well-known non-relativistic linear and rotational Doppler effect[21-24], respectively. Nevertheless, generally, directly extracting Doppler shift by optical interference cannot distinguish Doppler up or down shift (blue or red shift) with respect to the optical frequency, resulting in the failure of direction detection, unless additionally using dual-frequency (or heterodyne) detection[17,25-27].

Moreover, the movement of objects usually contains multiple dimensionalities. One of the most representative multi-dimensional movement is the helical motion, consisting of translational and rotational velocity vectors. This kind of movement universally exists in both natural world and human industrial society, for instances, the spirilla locomotion[28], the celestial helix motion, and some occasions of machine tool processing and space docking, etc. It is of great significance for simultaneous determination of both the translational and rotational velocities in some relevant researches and applications with this multi-dimensional movement. The commonplace techniques to get this multi-dimensional measurement refer to the means of multipath interferometric configurations[9], 3D dynamic imaging systems[29], and step-by-step detection or based on structured light with spatial polarization[30,31]. However, these schemes encounter expensive equipment, cumbersome system, complicated implementation, and/or failure to get absolute motion direction.

Here we demonstrate new structured light interferometry to achieve simultaneous determination of all motion information of this multi-dimensional movement, including the magnitude and direction information of both translational and rotational components. Beyond the simplex phase DoF of light, here we fully take advantage of the spatial amplitude, phase, and polarization DoFs of structured light. Such interferometric scheme can thoroughly break the fundamental limits of traditional light interferometry to get the simultaneous determination of multi-dimensional motion vectors. To make the concept on structured light interferometry better generality and integrity, we present two metrology scenarios in terms of small particles and bulk targets, respectively. At the basic framework of utilization of multiple Doppler shifts of structured light, two concrete interferometric schemes are individually conducted to experimentally retrieve the multi-dimensional velocity vectors for these two metrology scenarios.

## Results

### Concept and Principle

In general, the Poynting vector of a light beam can be written as $\mathbf{k}(k_x, k_y, k_z)$ in three-dimensional Cartesian coordinate. When interacting with a moving object with three-dimensional velocity vectors $\mathbf{v}(v_x, v_y, v_z)$, it gives the resulting Doppler shift $\Delta\omega = \mathbf{k} \cdot \mathbf{v} = k_x v_x + k_y v_y + k_z v_z$ that contains three contributions. It signifies that the multi-dimensional velocity can be determined provided that multiple Doppler shifts are available to solve the linear equations in multiple unknowns ($\Delta\omega_1$, $\Delta\omega_2$, and $\Delta\omega_3$). Nevertheless, as for the classically monochromatic plane light, its Poynting vector $\mathbf{k}$ is constant in space, giving the wavenumber $\sqrt{k_x^2 + k_y^2 + k_z^2} = 2\pi/\lambda$ with $\lambda$ being wavelength. Obviously, the traditional optical interferometry using monochromatic plane light just allows detection of one-dimensional movement along/against the direction of light beam.

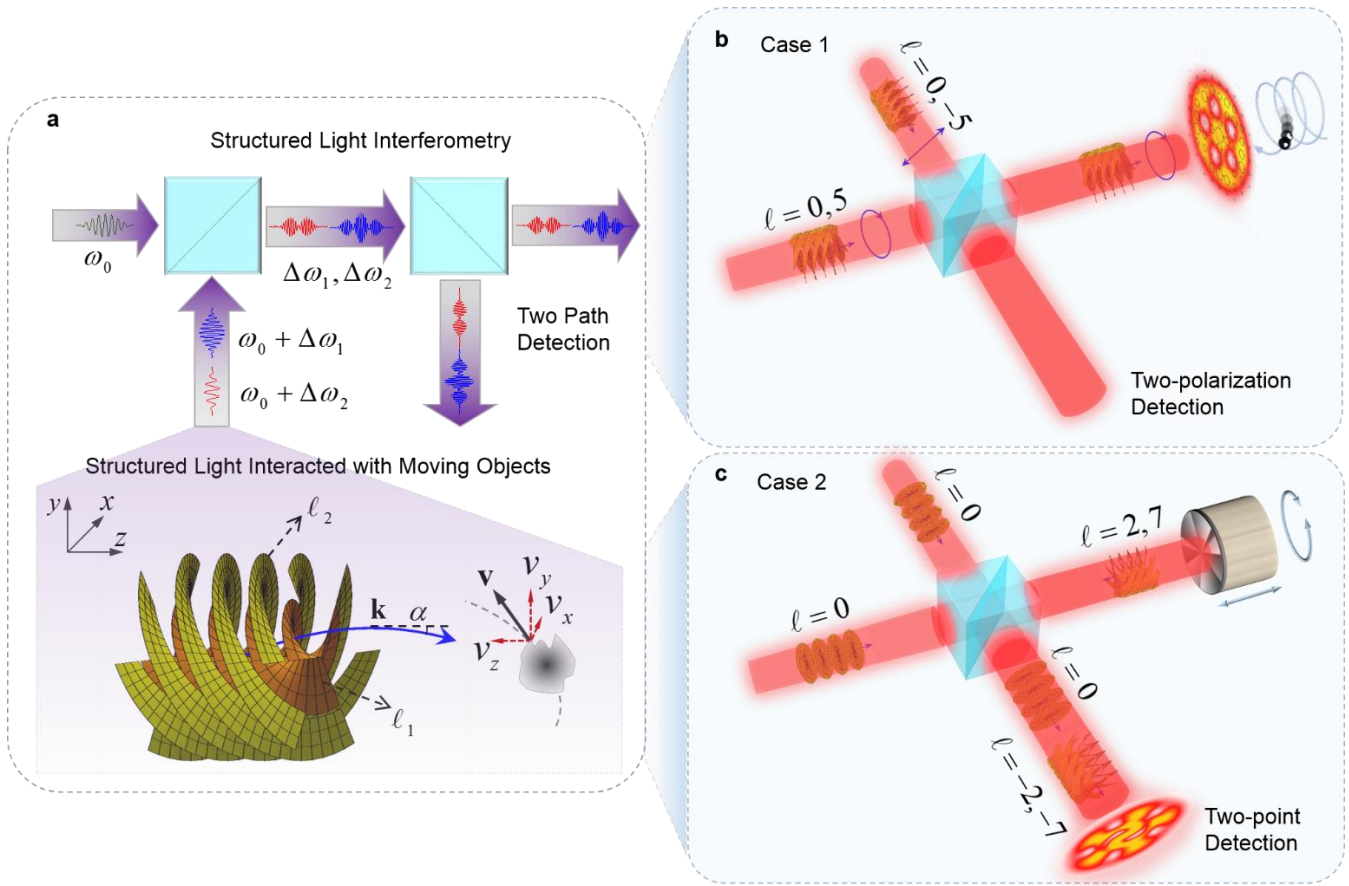

**Fig. 1 | Concept and schematic of structured light interferometry. a** Multiple Doppler shifts with different amplitudes are induced by interaction of the spatial-varying Poynting vectors of structured light with multi-dimensional vectors of moving objects. These Doppler shifts are extracted by interference and two path detection to retrieve the multi-dimensional movement. **b** The structured light ($\ell = 0, 5$) is used for detecting small particles by exploiting spatial amplitude and polarization DoFs of structured light. **c** The structured light ($\ell = 2, 7$) is used for detecting bulk targets by exploiting spatial intensity and structure of interference pattern.

As is well known, the twisted beam has a skew Poynting vector $\mathbf{k} = (-\ell\sin\phi/r, \ell\cos\phi/r, k_z)$ for light propagating along the $+z$ direction, where $\ell$ denotes the winding number, and $(r,\phi)$ represents the radial and azimuthal positions. This kind of Poynting vector features an inherent skew angle $\alpha = \tan^{-1}\left(\sqrt{k_x^2+k_y^2}/k_z\right) \approx \ell/kr$ with respect to the axis of light[32,33], where $k_z \simeq k$. In recent years, such structured light is gaining wide interests and may find myriad applications in manipulation, microscopy, metrology, astronomy, quantum processing, and optical communications[34-43]. By contrast with the classical plane light, significantly, the spatial-varying Poynting vector of structured light possesses more spatial dimensionalities that allow direct detection of a multi-dimensional movement $\mathbf{v} = (-\sin\phi \cdot r\Omega, \cos\phi \cdot r\Omega, v_z)$, where $\Omega$ is the rotational velocity. When interacted with this movement, such twisted light gives a Doppler shift $\Delta\omega = \mathbf{k}\cdot\mathbf{v} + kv_z = \ell\Omega + 2kv_z$. Note that the translational contribution of Doppler shift has the factor of 2, because of undergoing twice linear Doppler shifts for the light illuminating and then reflecting off the moving objects.

In this article we focus on the simultaneous determination of this multi-dimensional movement consist of velocity vectors $v_z$ and $\Omega$ based on the structured light (Fig. 1). Here a generally structured light beam with two twisted components ($\ell_1$ and $\ell_2$) is employed to get at least two Doppler shifts, such as $\Delta\omega_1 = 2kv_z + \ell_1\Omega$ and $\Delta\omega_2 = 2kv_z + \ell_2\Omega$ (or $\Delta\omega_3 = \Delta\omega_2 - \Delta\omega_1 = (\ell_2 - \ell_1)\Omega$), which serves as linear equations in two unknowns solved to directly retrieve the multi-dimensional velocities ($v_z$ and $\Omega$) of moving objects. Despite the means of multipath interferometric configurations (changing $\mathbf{k}$) and using structured light with step-by-step detection or based on spatially polarized states probably enable the multi-dimensional measurement, these schemes could greatly complicate the measuring system and implementation, and/or even still incapable of determining the absolute directions of velocity vectors[30,31].

Generally speaking, the detectable moving objects using optical interferometry can be mainly classified into two categories in terms of small particles and bulk targets. The approach of extracting Doppler shifts from structure light has distinct differences between these two different categories. Because the small particles locally reflect or scatter the light from structured fields, while the bulk targets usually reflect (scatter) off the overall structured light. Thus, without loss of generality, here we demonstrate two interferometric schemes to individually retrieve the multi-dimensional movement of small particles (Fig. 1b) and bulk targets (Fig. 1c).

Notably, to achieve the goal of multi-dimensional measurement, two key issues that have to be addressed in the experiment. One is how to recognize different Doppler shifts in Doppler frequency spectra. Here we take the method of distinguishing the relative amplitudes of Doppler shift peaks to recognize them. In principle, the amplitude DoF (power ratio) of twisted components of structured light can be controlled when interacting with the moving objects. Such interaction in different intensities may give the amplitude-distinguishable Doppler shift peaks. Another key issue is how to distinguish the signs of these Doppler shifts. Because when directly extracted by interference with reference light, the Doppler shifts usually reduce to the scalar values $|\Delta \omega|$ as experimental observables, due to the only intensity-sensitivity for most photodetectors, which results in the direction ambiguity of measurement. Here instead of using the well-known dual-frequency (or heterodyne) detection[17,25-27], we demonstrate novel methods to distinguish the signs of Doppler shifts by directly exploiting the spatial structure and polarization DoF of structured light. As for individual measurement of small particles and bulk targets, the means of orthogonal polarization detection (Fig. 1b) and the two-point detection (Fig. 1c) are implemented, respectively (see Methods).

**Experiment and Data**

**Case one: small particle**

The key part of interferometric configuration is shown for detecting a moving small particle (Fig. 2a). A structured beam ($\ell = 0, 5$ and $\lambda = 632.8$ nm) was beforehand generated by a spatial light modulator (SLM) uploaded with a complex phase hologram (see supplementary materials). This structure light was split into two paths. The measuring light was controlled as the circular state of polarization (SoP) before illuminating on the moving particle, while its mirrored one ($\ell = 0, -5$) as reference light is to the diagonal SoP before interference. In the experiment, the particle and its rotational velocity were mimicked by a digital micromirror device (DMD), and the translational velocity was gotten by putting DMD at an electrically driven sliding guide (SG) (see Methods).

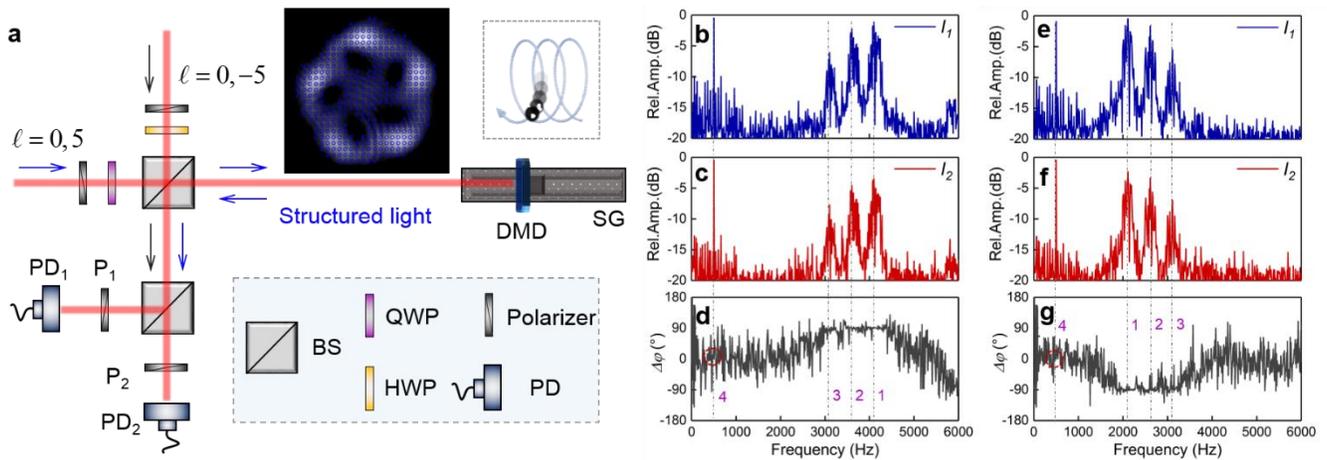

**Fig. 2 | Experimental configuration and measured results for the case of detecting a small particle. a** Sketch outline of structured light interferometry that allows for simultaneous measurement of multi-dimensional movement of a moving particle. Blue arrows denote a measuring light beam, and black arrows denote a reference light beam. The inset shows the structured light with circular SoP generated experimentally. BS: beam splitter; QWP: quarter wave plate; HWP: half wave plate; PD: photoelectric detector, SG: sliding guide. **b-d** Measured results of multi-dimensional movement with $v_z = 1$ μm/s and $\Omega = 100 \times 2\pi$ rad/s. **e-g** Measured results of multi-dimensional movement with $v_z = -1$ μm/s and $\Omega = 100 \times 2\pi$ rad/s. **b, e** Doppler

frequency-amplitude spectra ($I_1$) in the state of x-polarization gotten by $P_1$ and $PD_1$. **c**, **f** Doppler frequency-amplitude spectra ($I_2$) in the state of y-polarization gotten by $P_2$ and $PD_2$. **d**, **g** Doppler frequency-phase spectra as RPDs between these two Doppler signals with orthogonal SoPs (**b** and **c**, or **e** and **f**).

After locally reflected back from the structured field with circular SoP by the moving particle, the light was interfered with the reference light with diagonal SoP. Two photoelectric detectors (PDs) were used for synchronous collection of the interference signals in orthogonal SoPs (Fig. 2a). These Doppler time-domain signals collected in two paths carry full motion information of the moving particle. The measured results for multi-dimensional movement in two different states were obtained by making Fast Fourier transformation (FFT) for the collected Doppler signals (Figs. 2b-2g). Note that, for convenience, the frequency-phase spectra (Figs. 2d and 2g) were plotted as relative phase differences (RPDs) between these two orthogonally polarized Doppler signals. Other measured results as contrast were provided for multi-dimensional movement with minus rotational velocity (see supplementary materials).

From these measured results, all the Doppler signals exhibit as four Doppler frequency peaks, and for each movement state, its orthogonally polarized Doppler signals show the same frequency peaks (Figs. 2b, 2c, 2e and 2f). However, noticeably, three of them give RPDs about $90°$ (Fig. 2d) or $-90°$ (Fig. 2g). They are dependent of the relative direction of multi-dimensional movement, which importantly can be available to indicate the signs of Doppler shifts (see Methods). Note that the RPD of the remaining peak always fixes near zero regardless of the relative direction, and thus can be first recognized (marked 4) when analyzing measured data. Besides, these Doppler shift peaks are characterized by different relative amplitudes. To get the amplitude-distinguishable peaks, when conducting the experiment, the size (scaling with $\sqrt{|\ell|}$) and power radio of two twisted components of structured light should be subtly controlled to match the rotation radius of the moving particle. Here we marked the highest peak (except for peak 4 already recognized) with Doppler shift 1 and the lowest peak with Doppler shift 3. All these marked peaks correspond to Doppler shifts $|\Delta f_j| = |\Delta \omega_j|/2\pi$, $j$ = 1, 2, 3, and 4, where $|\Delta \omega_1| = |10\Omega + 2kv_z|$, $|\Delta \omega_2| = |5\Omega + 2kv_z|$, $|\Delta \omega_3| = |2kv_z|$, and $|\Delta \omega_4| = |5\Omega|$. Note that in this case the rotational contribution of Doppler shift $|\Delta \omega_1|$ not only depends upon the twisted component ($\ell = 5$) from the measuring structured light, but also the twisted component ($\ell = -5$) from the reference structured light (see supplementary materials). Thus, the velocity components ($v_z$ and $\Omega$) of multi-dimensional movement can be completely determined by solving the linear equations in two unknowns, for example, selecting the well distinguished Doppler shifts $\Delta f_1$ and $\Delta f_2$.

In general, this kind of multi-dimensional movement contains four types of helical motion states (Fig. 3a), of which the chirality is associated with the relative direction between translational and rotational velocity vectors. Based on the distinguished Doppler shifts above, we successfully retrieved the multi-dimensional velocities with four kinds of linear variation (Fig. 3b). The movement states in four quadrants of the rectangular coordinate correspond to four types of helical motion states (Fig. 3a). We presented the measured Doppler shifts and their RPDs versus the translational velocities for one case (Fig. 3c and 3d). The completely measured results for all cases were given (see supplementary materials). Here the experimental data comes from multiple measurements for each movement state. These repeated data, on the one hand, were used for tolerance analysis, and on the other hand to remove noise frequencies so as to better identify the targeted Doppler frequencies (see supplementary materials). In fact, the retrieved rotational velocities were rectified through the Doppler shift ($\Delta f_4$), because it was nearly undisturbed by the ambient (see supplementary materials). Especially, the RPDs exhibit as being adverse with the linear trend of Doppler shifts, which shows the feasibility of distinguishing the signs of Doppler shifts based on the RPDs.

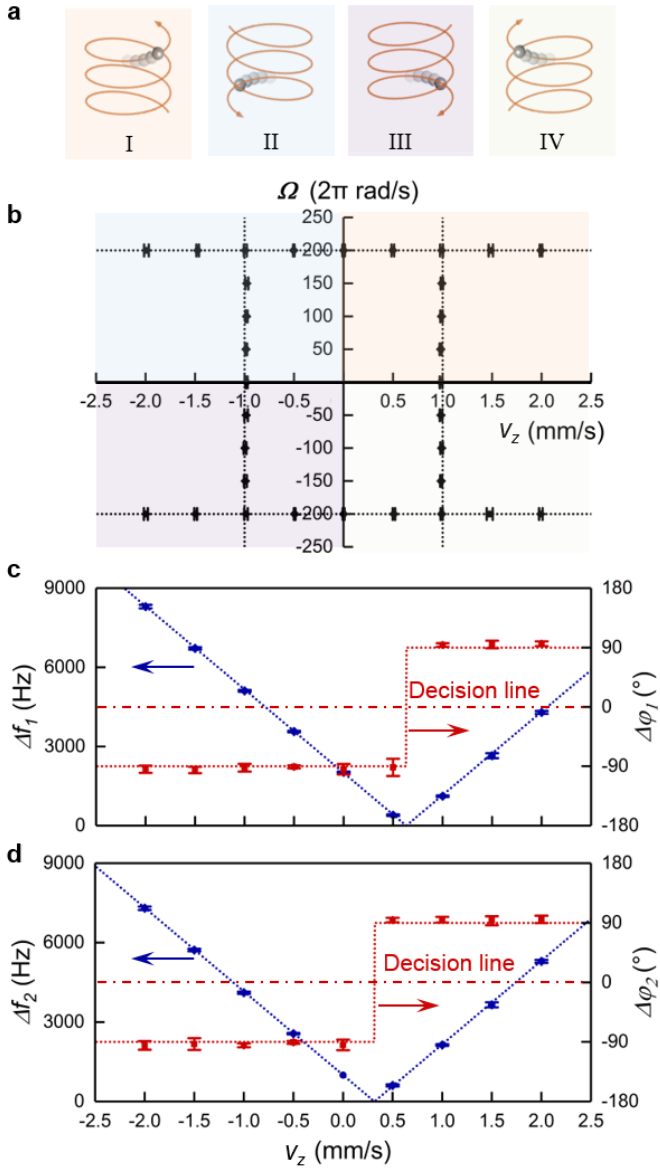

**Fig. 3 | Measured results under different states of multi-dimensional movement. a** Four different helical motion states of a moving small particle. **b** The experimentally retrieved multi-dimensional velocities expressed in rectangular coordinate where four quadrants correspond to four helical motion states (**a**). **c** Measured Doppler shifts $|\Delta f_1|$ and RPDs $\Delta\varphi_1$, **d** Measured Doppler shifts $|\Delta f_2|$ and RPDs $\Delta\varphi_2$, under the fixed rotational velocity $\Omega = -400\pi$ rad/s. The measured RPDs above the decision line indicate the plus signs of Doppler shifts, vice versa.

**Case two: bulk target**

In the standard interferometric configuration, a mirror needs to be installed on a translational bulk target to give a linear Doppler shift for the reflected measuring light. Naturally, it is completely reasonable to replace this mirror with a reflective phase mask (or a holographic pattern) to generate a structured light beam with multiple Doppler shifts, and thus predictably retrieve both the translational and rotational component of the moving target. Here we also experimentally demonstrate this simultaneously multi-dimensional measurement of a moving bulk target based on the structured light.

In this experiment, we employed the DMD holographic pattern to replace an equivalent complex phase mask installed on a moving bulk target that generates and meanwhile rotates the structured light. The DMD refers to the holography method[44] when generating structured light through both spatial amplitude and phase modulation (see supplementary materials). Analogous to Case one, the rotational velocity was mimicked by changing the time interval to continuously switch DMD holographic patterns, and the translational velocity was also controlled by the electrically driven sliding guide where the DMD was put.

The key part of interferometric configuration is shown (Fig. 4a). A Gaussian beam ($\lambda$=632.8 nm) was beforehand split into the measuring and reference light, respectively. In the measuring path, a designed DMD holographic pattern shaped the Gaussian light into the structured light ($\ell = 2, 7$) (see the left inset in Fig. 4a). An interference pattern was produced by suitably adjusting the power and size of the reference light relative to the measuring structured light (see the right inset in Fig. 4a). This structured interference pattern features two regions, the helical fringes in inner region and the petal-like fringes in outer region (see Fig. 4a and Methods). In principle, the complex phase mask (or a holographic pattern) can deliver full motion information of the moving target to the generated structured light, which manifests as the interference fringes rotating with different velocities in different regions (see supplementary materials). Their rotational directions depend upon the relative direction of multi-dimensional movement associated with the signs of Doppler shifts (see Methods). To get the Doppler shifts and meanwhile distinguish the signs of them, two PDs with a relative azimuthal angle were positioned at the junction of two interference regions. Additionally, to make these Doppler shifts amplitude-distinguishable, the radial positions of two PDs should also be adjusted in consideration of the relative power between two twisted components of the structured light.

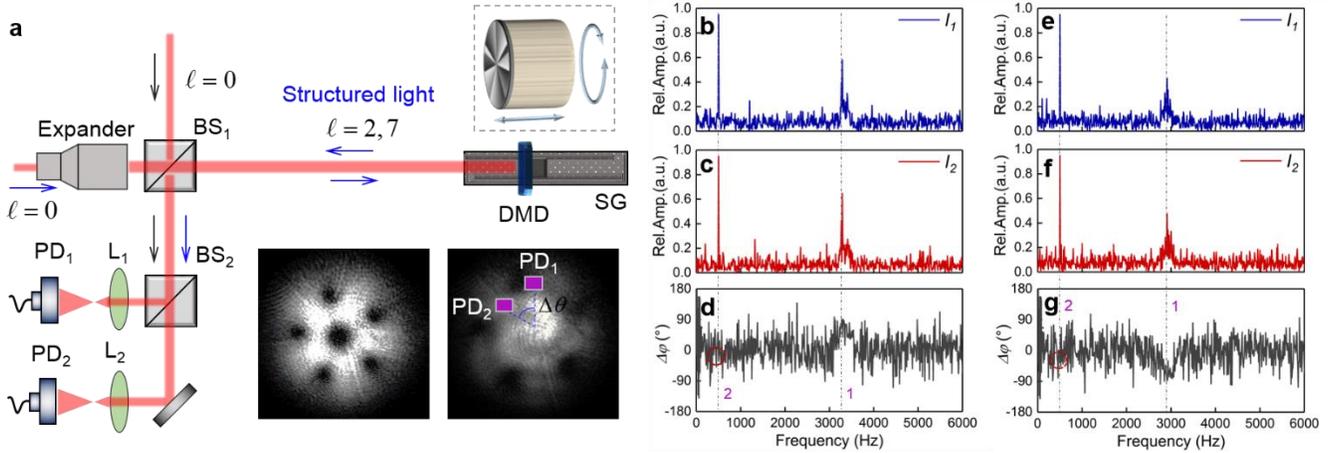

**Fig. 4 | Experimental configuration and measured results for the case of measuring a moving bulk target. a** Sketch outline of the structured light interferometer. BS: beam splitter; L: lens; PD: photoelectric detector. The insets show the structured light generated experimentally and its interference pattern with reference Gaussian beam. $PD_1$ and $PD_2$ were positioned at the junction of two interference regions with a relative azimuthal angle. **b-d** Measured results of multi-dimensional movement with $v_z = 1\,\mu m/s$ and $\Omega = 100 \times 2\pi$ rad/s. **e-g** Measured results of multi-dimensional movement with $v_z = -1\,\mu m/s$ and $\Omega = 100 \times 2\pi$ rad/s. **b**, **e** Doppler frequency-amplitude spectra gotten by $L_1$ and $PD_1$. **c**, **f** Doppler frequency-amplitude spectra gotten by $L_1$ and $PD_2$. **d**, **g** Doppler frequency-phase spectra as RPDs between two Doppler signals at the relative azimuthal positions (**b** and **c**, **e** and **f**).

The measured results of multi-dimensional movement in two states were presented (Figs. 4b-4g). Here we marked the low peak with Doppler shift 1 and the high peak with Doppler shift 2, corresponding to the observable Doppler shifts $|\Delta f_j| = |\Delta \omega_j|/2\pi$, $j=1$ and 2, where $|\Delta \omega_1| = 2|\Omega + kv_z|$ and $|\Delta \omega_2| = |5\Omega|$. Similar to Case one, the frequency-phase spectra (Figs. 4d and 4g) were also given as RPDs between two

Doppler signals collected in two relative positions (Figs. 4b and 4c, Figs. 4e and 4f). Other measured results as contrast were provided for the multi-dimensional movement with minus rotational velocity (see supplementary materials). From these measured results, the Doppler peak 1 shows a fixed RPD around $80°$ (Fig. 4d) or $-80°$ (Fig. 4g), and the Doppler shift 2 gives the fixed values (about $-35°$). These direction-dependent RPDs also can be available to distinguish the signs of Doppler shifts.

Similar to Case one, we show four different states of multi-dimensional movement (Fig. 5a), corresponding to four helical motion states (Fig. 3a) for an off-axis position at the moving bulk target. The multi-dimensional velocities with four kinds of linear variation (Fig. 5b) were successfully retrieved by solving the linear equations in two unknowns ($\Delta f_1$ and $\Delta f_2$). The measured Doppler shifts and their RPDs for one case were presented (Fig. 5c and 5d), as well as the complete measured data (see supplementary materials). All the measured Doppler shifts ($|\Delta f_1|$ and $|\Delta f_2|$) are in agreement with the theoretical analysis. Note that the measured RPDs, especially for $\Delta\varphi_1$ (around $\pm 70°$), show large errors, and even do not accord with the prediction (see Methods). Nonetheless, these errors actually do not affect the distinguishment of the signs of Doppler shifts, because of the large tolerance range (see supplementary materials).

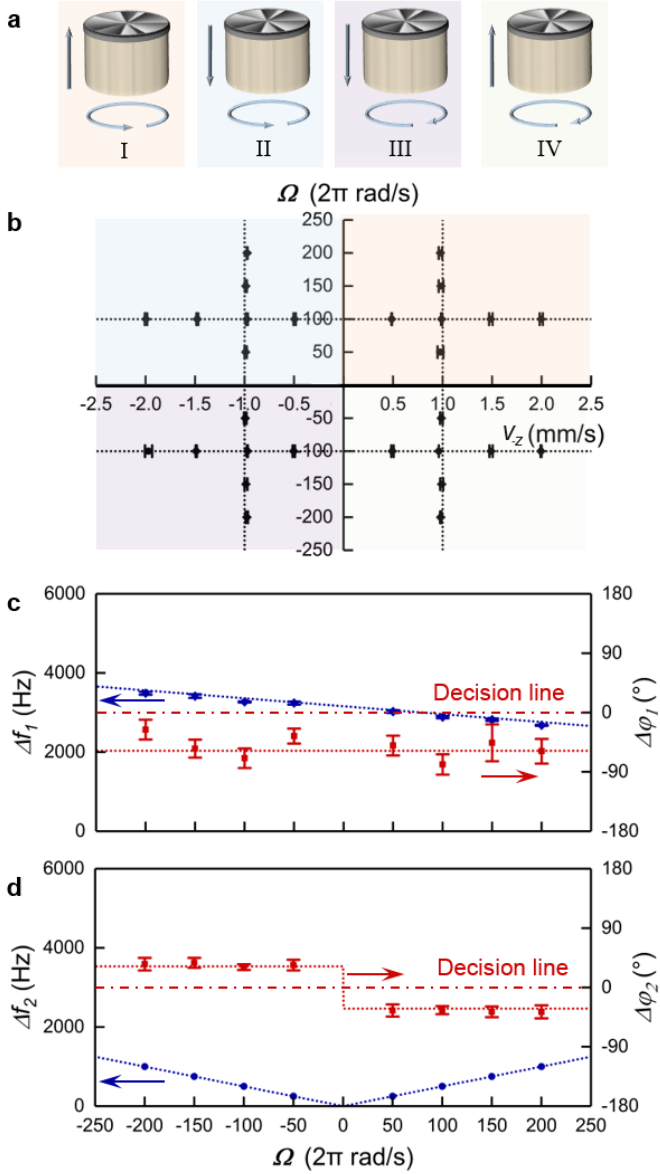

**Fig. 5 | Measured results under different states of multi-dimensional movement. a** Four different multi-dimensional movement states of a moving bulk target. **b** The experimentally retrieved multi-dimensional velocities expressed in rectangular coordinate where four quadrants correspond to four cases (**a**). **c** Measured Doppler shifts $|\Delta f_1|$ and RPDs $\Delta \varphi_1$, **d** Measured Doppler shifts $|\Delta f_2|$ and RPDs $\Delta \varphi_2$, under the fixed translational velocity $v = -1$ mm/s. In this case, the measured RPD $\Delta \varphi_1$ above/below the decision line indicates the plus/minus signs of Doppler shift $\Delta f_1$, while this judgement is reverse for $\Delta \varphi_2$ and $\Delta f_2$.

**Discussion**

We have experimentally demonstrated two individual interferometric schemes to successfully retrieve multi-dimensional move vectors for both small particles and bulk targets. For more generality, another interferometric scheme also based on structured light is conceptually shown to retrieve multi-dimensional of the general rough surfaces (see supplementary materials). In principle, the detectably multi-dimensional movement of moving objects, should be on condition that the translational direction is nearly parallel with the normal of the rotating plane of the moving objects. Additionally, when measuring the bulk targets where a complex phase mask needs to be installed, the central normal of the mask should be kept collinear with the rotation axis (see supplementary materials). Nevertheless, in the experiment here simulated by DMD, this restriction condition is unnecessary provided that the reflection direction of the diffraction order as the monitored light is adjusted to always parallel with the translational direction of DMD on the sliding guide (see supplementary materials). Despite this unlimited condition, all the experimental results here provide full verification of the feasibility of structured light interferometry used for simultaneous measurement of multi-dimensional motion vectors.

Such multi-dimensional simultaneous measurement can be attributed to the fully exploited multiple physical dimensionalities of structured light that give multiply amplitude- and direction-distinguishable Doppler shifts. Particularly, the detectability of absolute translational and rotational directions benefits from the RPDs between different Doppler signal components that come from two orthogonal SoPs (when detecting small particles) or two relative detection positions (when measuring bulk targets). Despite of some adoptable schemes using multipath interferometric configurations or other schemes using structured fields perhaps, the structured light interferometry here demonstrates more advantages to greatly simplify the implementation for the simultaneous measurement of multi-dimensional motion vectors.

**Table 1. Summary of feature comparison among typically interferometric schemes**

| Movement<br><br>Schemes based on | Translation alone | | Rotation alone | | Translation and Rotation | | | |
|---|---|---|---|---|---|---|---|---|
| | $|v|$ | Sign | $|\Omega|$ | Sign | $|v|$ | Sign | $|\Omega|$ | Sign |
| Plane-phase light | √ | √[#] | × | × | √ | √[#] | × | × |
| Single-twisted light | √ | √[#] | √ | √[#] | P | P | P | P |
| Superposed light with $\pm\ell$ [*] | × | × | √ | √[#] | × | × | √ | √[#] |
| Generally structured light | √ | √ | √ | √ | √ | √ | √ | √ |

Marking 'P': pending; Superscript '*': without a reference light; Superscript '#': special technique needed

Finally, we summarize several typically interferometric schemes based on plane-phase light, single twisted light, superposed light with opposite winding numbers, and the generally structured light (see Tab. 1 and see supplementary materials). The conventional scheme using the classical plane-phase light or superposed light with opposite winding numbers just allows the corresponding translational or rotational movement measurement. It should be noted that to get the direction information of movement, some special techniques need to be taken, for example, the orthogonal polarization and two point detection proposed here, or the traditional dual-frequency (or heterodyne) detection. As for the scheme using single twisted light, although possessing the ability of detecting a translation or rotation alone movement, such scheme fails to differentiate the translational and rotational components from the multi-dimensional movement. By contrast, the new interferometric interferometry base on generally structured light shows the remarkable function enhancement that allows the retrievement of all motion dimensionalities. Our theoretical and experimental results definitely demonstrate that the structured light interferometry can offer abundant Doppler information and enable expand optical interferometry to multi-parametric or multi-dimensional simultaneous metrology.

## Methods

### Distinguishing the signs (red or blue shift) of Doppler shifts

**Case one: small particle.** The extracted Doppler shifts ($\Delta\omega_j$, $j$ =1, 2, and 3) refer to interference between the scattered/reflected light from structured field with circular SoP ($\sigma$) by the moving particle and the reference light with diagonal SoP ($\tau$). Here we reveal an additional Doppler information about relative phase evolution between Doppler frequency shifts in two orthogonal SoPs. For convenience, we assume that each of these Doppler shifts derives from the measuring structured light, not related to the reference light. Its time-domain signal component collected can be given as

$$\mathbf{E}_j = \frac{1}{\sqrt{2}}\left(\mathbf{e}_x + i\sigma\mathbf{e}_y\right)\cdot\exp\left\{-i\left[\left(\omega+\Delta\omega_j\right)t+\beta_\ell\right]\right\}, \tag{1}$$

where $\sigma = +1$ or $-1$ denotes the right-handed or left-handed circular SoP, respectively. The reference light is

$$\mathbf{E}_{\text{ref}} = \frac{1}{\sqrt{2}}\left(\mathbf{e}_x + \tau\mathbf{e}_y\right)\cdot\exp\left[i\left(-\omega t+\beta_0\right)\right], \tag{2}$$

where $\tau = +1$ or $-1$ denotes the 45° or 135° diagonal SoP, respectively. The interference signal component between them can be deduced as

$$I_j = \left(\mathbf{E}_{\text{ref}} + \mathbf{E}_j\right) \cdot \left(\mathbf{E}_{\text{ref}} + \mathbf{E}_j\right)^* = \left[1 + \cos\left(\Delta\omega_j t - \Delta\beta\right)\right]\mathbf{e}_x + \left[1 + \sigma\tau\sin\left(\Delta\omega_j t - \Delta\beta\right)\right]\mathbf{e}_y$$

$$= \begin{cases} \left[1 + \cos\left(\Delta\omega_j t - \Delta\beta\right)\right]\mathbf{e}_x + \left[1 + \cos\left(\Delta\omega_j t - \sigma\tau\dfrac{\pi}{2} - \Delta\beta\right)\right]\mathbf{e}_y, & \Delta\omega_j > 0 \\ \left[1 + \cos\left(\Delta\omega_j t - \Delta\beta\right)\right]\mathbf{e}_x + \left[1 + \cos\left(\Delta\omega_j t + \sigma\tau\dfrac{\pi}{2} - \Delta\beta\right)\right]\mathbf{e}_y, & \Delta\omega_j < 0 \end{cases}, \quad (3)$$

where $\Delta\beta = \beta_\ell - \beta_0$ is the phase difference between the collected measuring and reference light components. Notably, there is a specific delay/advance in time ($\Delta t = \pm\sigma\tau\pi/2$) between the x- and y-polarized Doppler signals, which is dependent of the signs of Doppler shifts $\Delta\omega_j$.

Such signal delay/advance originates form the relative velocity between translational ($v_z$) and rotational ($r\Omega$) vector components of the moving particle with respective to the skew Poynting vector of structured light (Fig. 1a). Because this moving particle can give different initial phase $\varphi$ when locally reflecting (scattering) the components with different linear polarization within the circularly polarized field (Fig. 6a). As for the collected orthogonal x- and y-polarized components, a resulting delay/advance in time may form between their Doppler time-domain signals for the particle moving with different relative translational and rotational velocities (Fig. 6c). For example, for $\Delta\omega_1$ associated with $\ell_1$, the delay (advance) is given by the relative velocity $v_z > -\alpha r\Omega$ ($v_z < -\alpha r\Omega$), where $\alpha = \ell_1/kr$. This delay/advance of Doppler signals can be embodied as a fixed RPD $\Delta\varphi = \varphi_y - \varphi_x$ between the orthogonally polarized Doppler shifts. In the experiment, such phenomenon was feasibly used for distinguishing the signs of Doppler shifts in Doppler frequency-phase spectra, as follows,

$$\text{sign}(\Delta\omega_j) = \begin{cases} +, & \Delta\varphi_j = \tau\sigma\pi/2 \\ -, & \Delta\varphi_j = -\tau\sigma\pi/2 \end{cases}, \quad (4)$$

where sign () represents the sign function. Note that the rotational contribution of Doppler shifts is actually determined by both the measuring and reference structured light. The true derivation of Doppler shifts and complete interference deduction were provided (see supplementary materials).

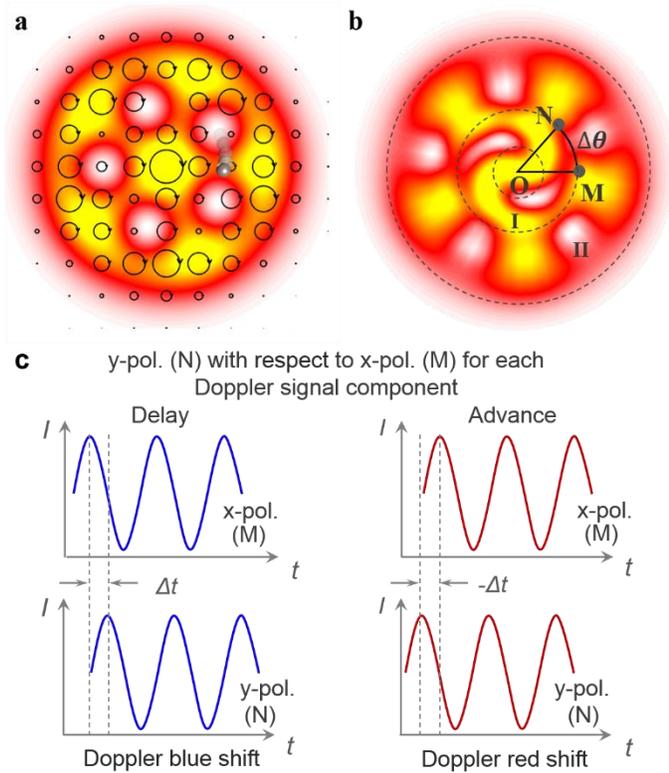

**Fig. 6 | Doppler signal delay/advance used for distinguishing the signs of Doppler shifts. a** The moving particle gives delay/advance in time for Doppler shift components ($\Delta\omega_j$, $j = 1, 2$, and $3$) between two orthogonal x- and y-polarized Doppler signals when reflected (scattered) from the structured light ($\ell = 0, 5$) with circular SoP. **b** The structured light ($\ell = 2, 7$) generated by a complex phase mask attached on a moving target features the rotating interference pattern with two regions. The relative azimuthal positions M and N are monitored to extract Doppler signals with delay/advance between M and N. **c** Each Doppler time-domain signal component for both cases (**a** and **b**) exhibits as delay/advance between light with x-polarization (monitored in M) and light with y-polarization (monitored in N) that can be available to distinguish the signs (blue or red shift) of Doppler shifts.

**Case two: bulk target.** The interference pattern between structured light ($\ell_1 = 2$, $\ell_2 = 7$) and reference light ($\ell = 0$) features two interference regions (Fig. 6b). The helical fringes in inner region I is produced by interference between the twisted component ($\ell_1 = +2$) and the reference beam; while the petal-like fringes in outer region II is as the superposition between two twisted components. If the complex phase mask is in the state of translational and rotational movement, these fringes in different regions will rotate with different velocities. The rotational velocity of helical fringes in inner region is $\Omega_1 = (\ell_1 \Omega + 2kv_z)/\ell_1$, and that of petal-like fringes in outer region is $\Omega_2 = \Omega$. When off-axis monitoring this rotating interference pattern at the junction of two interference regions, its inner and outer regions give the Doppler shift components $\Delta\omega_1 = 2kv_z + \ell_1 \Omega$ and $\Delta\omega_2 = |\ell_2 - \ell_1|\Omega$, respectively (see supplementary materials).

The rotational velocities $\Omega_1$ and $\Omega_2$ of interference pattern are directional, i.e., clockwise or counterclockwise rotation, which depends upon the relative direction of moving bulk targets. For example, for $\Omega_1$, the clockwise (counterclockwise) rotation is determined by the relative velocity $v_z > -\alpha r\Omega/2$ ($v_z < -\alpha r\Omega/2$), where $\alpha = \ell_1/kr$. So if two positions M and N with a relative azimuthal angle $\Delta\theta$ are selected as monitored points (Fig. 6b), the interference fringes rotating between M and N may produce relative delay/advance for the collected Doppler signals, which can be deduced by

$$\Delta t_j = \Delta\theta/\Omega_j \pm 2n\pi/|\Delta\omega_j|, \tag{5}$$

where $n = 0, 1, 2\ldots$, $j = 1$ and 2 indicates the Doppler signal components extracted from the inner and outer interference regions, respectively. Similar to the Case one, such delay/advance can give a fixed RPD between Doppler shifts collected in these two relative positions, that is,

$$\Delta\varphi_j = |\Delta\omega_j|\Delta t_j = \Delta\ell_j \Delta\theta \cdot \text{sign}(\Delta\omega_j) \pm 2n\pi, \tag{6}$$

where $\Delta\ell_1 = \ell_1$, $\Delta\ell_2 = |\ell_2 - \ell_1|$. In the experiment, similarly, these RPDs were used for distinguishing the signs of Doppler shifts in Doppler frequency-phase spectra, as follows,

$$\text{sign}(\Delta\omega_j) = \begin{cases} +, & \Delta\varphi_j = \Delta\ell_j \Delta\theta \\ -, & \Delta\varphi_j = -\Delta\ell_j \Delta\theta \end{cases}, \quad (7)$$

For example, in this experiment, $\Delta\theta = -7°$, $\ell_1 = -2$, and $\ell_2 = -7$, the signs of Doppler shift $\Delta\omega_1$ can be distinguished in this way: $\Delta\varphi_2 = 14°$ for blue shift ($\Delta\omega_1 > 0$, or $v_z > -\alpha r'\Omega/2$), while $\Delta\varphi_1 = -14°$ for red shift ($\Delta\omega_1 < 0$, or $v_z < -\alpha r'\Omega/2$). The signs of Doppler shift $\Delta\omega_2$ can be distinguished as: $\Delta\varphi_2 = 35°$ for blue shift ($\Delta\omega_2 < 0$, or $\Omega < 0$), while $\Delta\varphi_2 = -35°$ for red shift ($\Delta\omega_2 > 0$, or $\Omega > 0$).

**Experimentally simulating and controlling the multi-dimensionally moving objects**

In the experiment, the moving particle was mimicked by a lump of 78 adjacent micromirrors of a digital micromirror device (DMD) in the 'ON' state. The complex phase mask installed on the moving bulk targets was also substituted by DMD. Instead of phase-only modulation through SLM, the generation of structured light using DMD refers to the holography method with both amplitude and phase modulation[44] (see supplementary materials). The rotational velocity for both the mimicked small particle and the DMD holographic patterns was controlled by changing the time interval to continuously switch the DMD patterns prepared beforehand. In addition, the translational velocity of moving objects was gotten by an electrically driven sliding guide where the DMD was putted. In the experiment, the position of the sliding guide should be subtly adjusted to keep the reflection direction of the diffraction order monitored as the targeted structured light always parallel with the translational direction of DMD on this sliding guide. When collecting data, the effective translation distance was about 1 mm for every measurement of each multi-dimensional movement state.

**Acknowledgments**

This work was supported by the National Natural Science Foundation of China (NSFC) under grants 11774116, 61761130082, 11574001, 11274131 and 61222502, the National Basic Research Program of China (973 Program) under grant 2014CB340004, the Royal Society-Newton Advanced Fellowship, the National Program for Support of Top-notch Young Professionals, and the Program for HUST Academic Frontier Youth Team. The authors acknowledge Prof. Miles J. Padgett at University of Glasgow for his technical supports and helpful discussions.

**Author contributions**

J.W. and L.F. developed the concept and conceived the experiments. Z.W. and L.F carried out the experiments and acquired the experimental data. L.F. performed the theoretical analyses. Z.W. and L.F. carried out the data analysis. L.F. contributed to writing the paper. J.W. finalized the paper. J.W. supervised the project.

**Supplementary Materials:**

Methods

Supplementary Text

Figures S1-S16

# Supplementary Materials for

## Structured light interferometry

Liang Fang, Zhenyu Wan, Jian Wang[*]

Wuhan National Laboratory for Optoelectronics, School of Optical and Electronic Information, Huazhong University of Science and Technology, Wuhan 430074, Hubei, China.

* Correspondence to: jwang@hust.edu.cn

# Case One: Detecting Small particles

## 1.1 Derivation of Doppler Effect Induced by Particles Moving in Structured Field

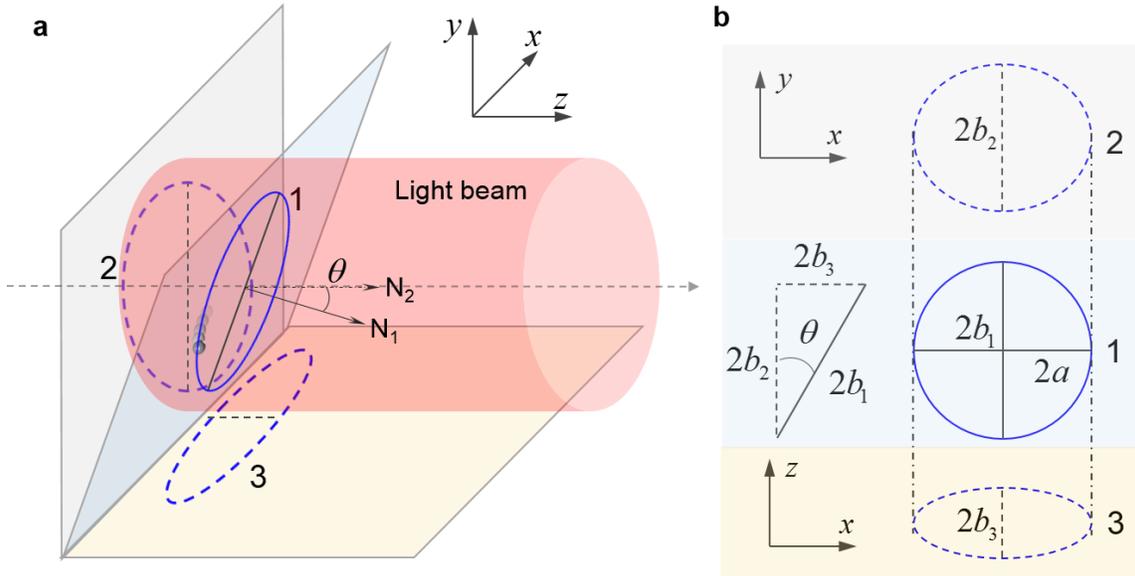

**Fig. S1 A moving particle in optical field rotates with an inclined angle $\theta$ with respect to the axis of light. a.** The circularly rotational trajectory1 could be projected as an ellipse2 on the transverse plane and an ellipse3 on the longitudinal plane. **b**. The relationship of long and short axes of these three trajectories.

As for detection of a moving particles in optical interferometry, the size of particle should be usually smaller than half period of interference fringes. The detected particles are considered to not have the property of birefringence so that keep the state of polarization when reflecting (or scattering) light based on the Mie's theory. Here to present analyses under both rotation and translation, without loss of generality, the rotation plane for a moving particle is arbitrary (Fig. S1), of which the normal $N_1$ in this case has an angle of $\theta$ with respect to the normal $N_2$ of transverse plane of light beam, namely, the axis of light.

In this case, the circular trajectory1 of rotation could be projected as ellipse2 and ellipse3 on the transverse plane and longitudinal plane, respectively. The long axes of both two ellipses are $2a$, the same as the diameter of circle1, while the short axis of ellipse2 is projected as $2b_2 = 2a \cdot \cos\theta$, and that of ellipse3 is $2b_3 = 2a \cdot \sin\theta$. The equation of trajectory1 can be given as $x_1 = a\cos\Omega t$ and $y_1 = a\sin\Omega t$, where $a$ is the radius of circular trajectory1, and $\Omega$ is the rotational velocity of a moving particle. Thus, when projecting into two ellipses, the corresponding equations can be written as $x_2 = x_1$ and $y_2 = \cos\theta \cdot y_1$ for the trajectory2, while $x_3 = x_1$ and $z_3 = \sin\theta \cdot y_1$ for the trajectory3.

Considering the translational velocity ($v_z$) nearly parallel to the axis of light beam, the trajectory of three-dimensional movement of the moving particle can be given by

$$\mathbf{s}(t) = a\cos\Omega t \cdot \vec{e}_x + a\cos\theta\sin\Omega t \cdot \vec{e}_y + (v_z t + a\sin\theta\sin\Omega t) \cdot \vec{e}_z, \tag{S1}$$

and its corresponding three-dimensionally instantaneous velocity,

$$\mathbf{v}(t) = \frac{d\vec{s}(t)}{dt} = -a\Omega\sin\Omega t \cdot \vec{e}_x + a\Omega\cos\theta\cos\Omega t \cdot \vec{e}_y + (v_z + a\Omega\sin\theta\cos\Omega t) \cdot \vec{e}_z. \tag{S2}$$

Here the sign (-/+) of $v_z$ and $\Omega$ is defined as the forward/backward translation and clockwise/counterclockwise rotation of moving objects respectively, with respect to the observation direction along the measuring light before illuminating on the moving object. If this optical field is a twisted light beam ($\ell$), its three-dimensional Poynting vector can be written as,

$$\mathbf{k} = -\frac{\ell}{r}\sin\phi \cdot \vec{e}_x + \frac{\ell}{r}\cos\phi \cdot \vec{e}_y + k \cdot \vec{e}_z, \tag{S3}$$

where $k = \omega/c$ is the wavenumber, $\omega$ is the angular frequency of light and $c$ is the light velocity in vacuum. When interacting with the moving particle, in the position of this particle, the space-time varying Poynting vector is

$$\mathbf{k}'(t) = -\frac{\ell}{r'(t)}\sin[\phi'(t)] \cdot \vec{e}_x + \frac{\ell}{r'(t)}\cos[\phi'(t)] \cdot \vec{e}_y + k \cdot \vec{e}_z, \tag{S4}$$

where the radial function is determined by the time-varying radii of ellipse2 across the transverse plane of light beam,

$$r'(t) = \sqrt{x_2^2 + y_2^2} = a\sqrt{\cos^2 \Omega t + \cos^2 \theta \sin^2 \Omega t}, \tag{S5}$$

and the azimuthal function is the time-varying angle of ellipse2,

$$\phi'(t) = \tan^{-1}\left(\frac{y_2}{x_2}\right) = \begin{cases} \tan^{-1}(\cos\theta \tan \Omega t) & 2n\pi - \frac{\pi}{2} \leq \Omega t < 2n\pi + \frac{\pi}{2} \\ \tan^{-1}(\cos\theta \tan \Omega t) + \pi & 2n\pi + \frac{\pi}{2} \leq \Omega t < 2n\pi + \frac{3\pi}{2} \end{cases}. \tag{S6}$$

Therefore, the instantaneous Doppler shift of reflected or scattered light by this moving particle from the twisted field can be expressed by

$$\Delta\omega = \mathbf{k}' \cdot \mathbf{v} + k(v_z + a\Omega \sin\theta \cos\Omega t)$$
$$= \frac{a\ell\Omega}{r'(t)}\sin[\phi'(t)]\sin\Omega t + \frac{a\ell\Omega}{r'(t)}\cos[\phi'(t)]\cos\theta \cos\Omega t + 2k(v_z + a\Omega \sin\theta \cos\Omega t). \tag{S7}$$

Using the relationships of $\tan^{-1}(x) = \sin^{-1}\left[x/\sqrt{1+x^2}\right] = \cos^{-1}\left[1/\sqrt{1+x^2}\right]$, Eq. (S7) can be simplified as

$$\Delta\omega = \frac{\cos\theta}{1+(\cos^2\theta - 1)\sin^2\Omega t}\ell\Omega + 2k(v_z + a\Omega \sin\theta \cos\Omega t). \tag{S8}$$

From another perspective, the multi-dimensional movement of the moving particle can be regarded as a comprehensive phase modulation for the reflected (scattered) light from the twisted component ($\ell$) of a structured light beam. Accordingly, the resulting phase modulation function should be expressed as

$$\Delta\Phi(t) = \ell\phi' + 2k\left(v_z t + a\sin\theta\sin\Omega t\right).$$

(S9)

It gives a time-varying Doppler shift $\Delta\omega = \partial\Phi(t)/\partial t$ with the same expression as Eq. (S8). Especially, under the movement state with coaxially rotational normal and translational direction, i.e., $\theta = 0$, this Doppler shift reduces to the well-known version,

$\Delta\omega = \ell\Omega + 2kv_z.$  (S10)

## 1.2 Extracting Multiple Doppler Shifts by Interference with Structured Reference Light

When this particle moves in a general structured field with two circularly polarized twisted components ($\ell_1$ and $\ell_2$), the reflected (scattered) light with phase modulation (S9) can be written as

$$\mathbf{E} = A_1(r)\cdot\exp\left\{-i\left[\omega t + \Delta\Phi(\ell_1,t)\right]\right\}\mathbf{e}_\sigma + A_2(r)\cdot\exp\left\{-i\left[\omega t + \Delta\Phi(\ell_2,t)\right]\right\}\mathbf{e}_\sigma,$$

(S11)

where $\mathbf{e}_\sigma = \left(\mathbf{e}_x + i\sigma\mathbf{e}_y\right)/\sqrt{2}$ indicates the state of circular polarization, and $\sigma = +1$ or $-1$ denotes right-handed or left-handed circular state of polarization (SoP), respectively. $A_1(r)$ and $A_2(r)$ denote the complex amplitudes of reflected light from twisted components $\ell_1$ and $\ell_2$, respectively.

To extract the Doppler shifts and RPDs, in general, we assume that the reference light is also a structured light with two twisted components ($\ell'_1$ and $\ell'_2$) but with diagonal SoP,

$$\mathbf{E}_{\text{ref}} = \left\{ A_1'(r) \cdot \exp\left[i(-\omega t + \ell_1'\phi)\right] + A_2'(r) \cdot \exp\left[i(-\omega t + \ell_2'\phi)\right] \right\} \mathbf{e}_\tau, \tag{S12}$$

where $\mathbf{e}_\tau = (\mathbf{e}_x + \tau \mathbf{e}_y)/\sqrt{2}$ describes the state of diagonal SoP, and $\tau = +1$ or $-1$ denotes 45° or 135° diagonal SoP, respectively. $A_1'(r)$ and $A_2'(r)$ indicate the complex amplitudes of these two twisted components.

The interference light between the reflected (scattered) light (S11) and the reference light (S12) can be deduced by

$$I = (\mathbf{E}_{\text{ref}} + \mathbf{E}) \cdot (\mathbf{E}_{\text{ref}} + \mathbf{E})^* =$$
$$= \frac{1}{2}\begin{Bmatrix} |A_1'|^2 + |A_2'|^2 + |A_1|^2 + |A_2|^2 \\ + A_1'A_2'\cos(\Delta\ell'\phi) + A_1'A_1\cos\left[\Delta\Phi(\ell_1,t) - \ell_1'\phi\right] \\ + A_1'A_2\cos\left[\Delta\Phi(\ell_2,t) - \ell_1'\phi\right] + A_2'A_1\cos\left[\Delta\Phi(\ell_1,t) - \ell_2'\phi\right] \\ + A_2'A_2\cos\left[\Delta\Phi(\ell_2,t) - \ell_2'\phi\right] + A_1A_2\cos\left[\Delta\Phi(\ell_2,t) - \Delta\Phi(\ell_1,t)\right] \end{Bmatrix}\mathbf{e}_x$$
$$+ \frac{1}{2}\begin{Bmatrix} |A_1'|^2 + |A_2'|^2 + |A_1|^2 + |A_2|^2 \\ + \tau^2 A_1'A_2'\cos(\Delta\ell'\phi) + \sigma\tau A_1'A_1\sin\left[\Delta\Phi(\ell_1,t) - \ell_1'\phi\right] \\ + \sigma\tau A_1'A_2\sin\left[\Delta\Phi(\ell_2,t) - \ell_1'\phi\right] + \sigma\tau A_2'A_1\sin\left[\Delta\Phi(\ell_1,t) - \ell_2'\phi\right] \\ + \sigma\tau A_2'A_2\sin\left[\Delta\Phi(\ell_2,t) - \ell_2'\phi\right] + \sigma^2 A_1A_2\cos\left[\Delta\Phi(\ell_2,t) - \Delta\Phi(\ell_1,t)\right] \end{Bmatrix}\mathbf{e}_y, \tag{S13}$$

where, $\Delta\ell' = \ell_2' - \ell_1'$. It is worth noting that when the reflected (scattered) light and reference structured light are synchronously collected by a group of lens and detectors, the interference positions on the photosurface of photoelectric detector (CCD) may rotate, following with the rotating signal light reflected (scattered) from the rotating particle. It gives an equivalent rotation of the local phase for the twisted components ($\ell_1'$ and $\ell_2'$) of the reference light. Accordingly, all the interference terms in (S13) that refer to the twisted components from the reference light should induce this phase rotation, that is, $\phi$ should be substituted by $\phi_0 + \phi'$, where $\phi_0$ is the initial phase, and $\phi'$ is given by Eq. (S6).

Therefore, for both the detected x- and y-polarized interference light, in principle, five different time-varying Doppler shifts can be extracted, as follows,

$$\Delta\omega_{\ell_1-\ell'_1} = \frac{\partial}{\partial t}\left[\Delta\Phi(\ell_1,t) - \ell'_1(\phi_0 + \phi')\right] = \delta(\ell_1 - \ell'_1)\Omega + 2k(v_z + \gamma) \ , \tag{S14}$$

$$\Delta\omega_{\ell_2-\ell'_1} = \frac{\partial}{\partial t}\left[\Delta\Phi(\ell_2,t) - \ell'_1(\phi_0 + \phi')\right] = \delta(\ell_2 - \ell'_1)\Omega + 2k(v_z + \gamma), \tag{S15}$$

$$\Delta\omega_{\ell_1-\ell'_2} = \frac{\partial}{\partial t}\left[\Delta\Phi(\ell_1,t) - \ell'_2(\phi_0 + \phi')\right] = \delta(\ell_1 - \ell'_2)\Omega + 2k(v_z + \gamma), \tag{S16}$$

$$\Delta\omega_{\ell_2-\ell'_2} = \frac{\partial}{\partial t}\left[\Delta\Phi(\ell_2,t) - \ell'_2(\phi_0 + \phi')\right] = \delta(\ell_2 - \ell'_2)\Omega + 2k(v_z + \gamma), \tag{S17}$$

$$\Delta\omega_{\ell_2-\ell_1} = \frac{\partial}{\partial t}\left[\Delta\Phi(\ell_2,t) - \Delta\Phi(\ell_1,t)\right] = \delta(\ell_2 - \ell_1)\Omega \ , \tag{S18}$$

where the correction factors $\delta = \cos\theta / \left[1 + (\cos^2\theta - 1)\sin^2\Omega t\right]$ and $\gamma = a\Omega\sin\theta\cos\Omega t$. Especially, in the case of angle $\theta \simeq 0$, the correction factor $\delta = 1$, and $\gamma = 0$, these time-varying Doppler shifts reduce to the well-known static Doppler shifts.

It should be stressed that the rotational contribution of these Doppler shifts is different from the familiar version that is usually not associated with the reference light without any twisted components. For example, the well-known Doppler shift $\Delta\omega = \ell\Omega + 2kv_z$, extracted by interference with reference Gaussian beam, its rotational contribution originates from the measuring structured light ($\ell$). In the main text, the measuring light is a structured field ($\ell_1 = 5$, $\ell_2 = 0$), while the reference light is its mirrored field ($\ell'_1 = -5$, $\ell'_2 = 0$), and $\theta \simeq 0$. So there are four observable Doppler shifts $|\Delta\omega_1| = |10\Omega + 2kv_z|$, $|\Delta\omega_2| = |5\Omega + 2kv_z|$, $|\Delta\omega_3| = |2kv_z|$, and $|\Delta\omega_4| = |5\Omega|$ in the Doppler frequency spectra.

## 1.3 Influence of The Inclined Rotation Plane on Extraction of Multiple Doppler Shifts

From these expressions (S14 to S18), the Doppler shifts usually undergo less perturbation from the former factor than the latter factor so that the factor $\delta$ can be neglected. It can be seen through the calculated results (Fig. S2-S4), where the Doppler shift $\Delta f_4 = \Delta \omega_4 / 2\pi$ is not affected by the inclined angles even though under a larger inclined angle. It means that the structured light interferometry can detect the rotational components of moving objects regardless of its inclined rotation plane. However, this inclined angle has severe influence on other three Doppler shifts that contain both rotational and translational contributions, which results from the longitudinal perturbation by the inclined rotation plane of the moving particle. Such influence could be neglected as long as such longitudinal perturbation caused by the inclined angle has an order of magnitude less than half of wavelength ( $2a\sin\theta < \lambda/2$ ). From the calculated results (Fig. S2-S4), if this inclined angle is larger than $0.01°$ (causing the perturbation length of 349 nm larger than half of wavelength 316.4 nm), it will be too difficult to distinguish the targeted Doppler shifts among a series of frequency peaks. Despite the measurement being so sensitive on this angle, in principle, the rotational and translational velocity components would also be determined, if the detectors were improved to have enough high response speed to get the nearly instantaneous Doppler shifts by extracting sequential ultra-short Doppler time-domain signals. Additionally, in this case, the inclined angle $\theta$ could also be determined by directly analyzing the alternating current component of the Doppler shifts under the already known rotation radius $a$.

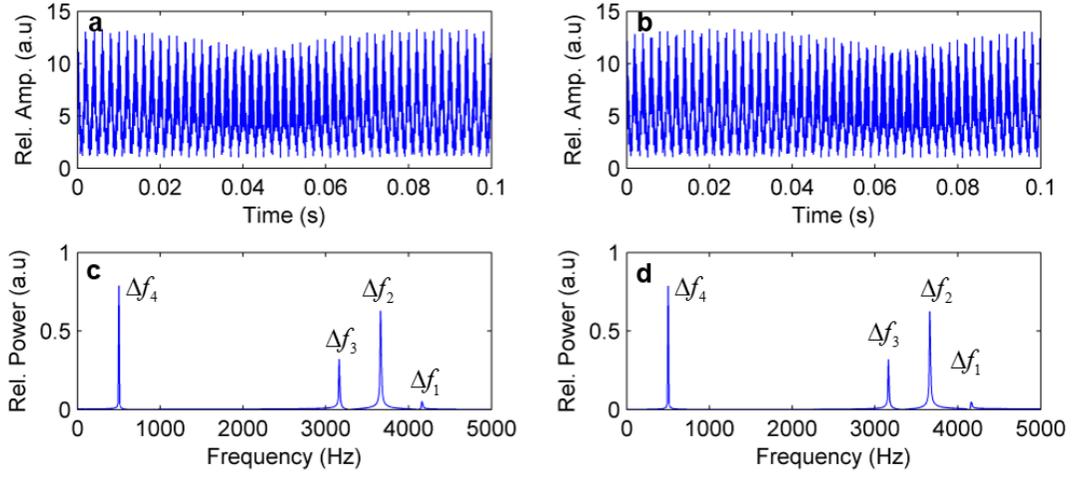

**Fig. S2.** Calculated results for multi-dimensional movement with $v = 1$ mm/s and $\Omega = 100 \times 2\pi$ rad/s, **under inclined angle** $\theta \simeq 0$, **rotation radius** $a = 1$ mm, **and** $\ell_1 = 5$, $\ell_2 = 0$. **a, b** Doppler time-varying signals for received x-polarized (**a**) and y-polarized (**b**) light. **c, d** Fourier frequency spectra for received x-polarized (**c**) and y-polarized (**d**) light.

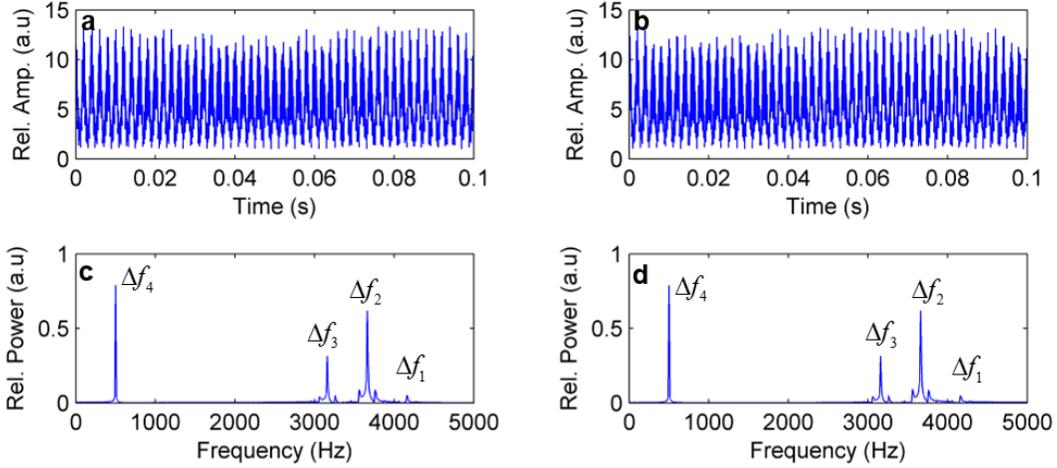

**Fig. S3.** Calculated results for multi-dimensional movement with $v = 1$ mm/s and $\Omega = 100 \times 2\pi$ rad/s, **under inclined angle** $\theta \simeq 0.001°$, **rotation radius** $a = 1$ mm, **and** $\ell_1 = 5$, $\ell_2 = 0$. **a, b** Doppler time-varying signals for received x-polarized (**a**) and y-polarized (**b**) light. **c, d** Fourier frequency spectra for received x-polarized (**c**) and y-polarized (**d**) light.

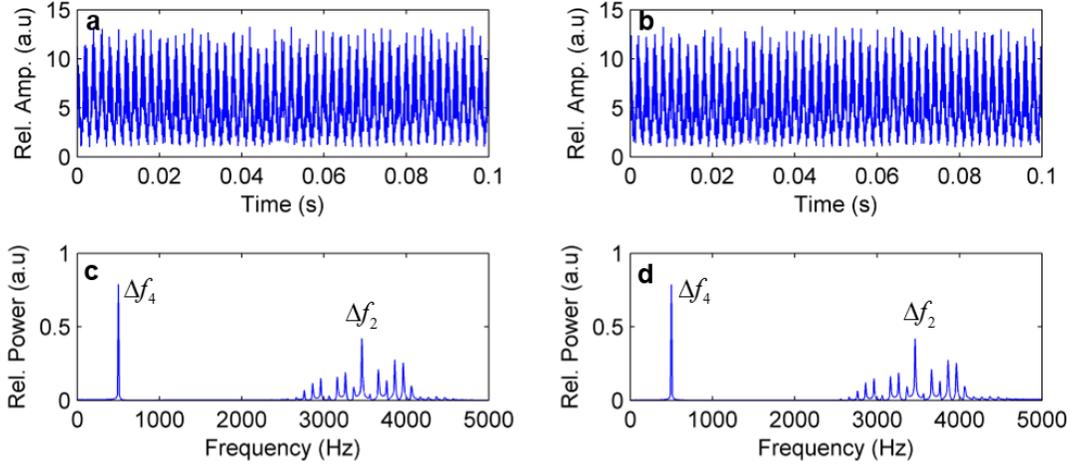

**Fig. S4. Calculated results for multi-dimensional movement with** $v=1\,\text{mm/s}$ **and** $\Omega = 100\times 2\pi\,\text{rad/s}$ **, under inclined angle** $\theta \simeq 0.01°$ **, rotation radius** $a=1\,\text{mm}$ **, and** $\ell_1 = 5$ **,** $\ell_2 = 0$ **. a, b** Doppler time-varying signals for received x-polarized (**a**) and y-polarized (**b**) ligt. **c, d** Fourier frequency spectra for received x-polarized (**c**) and y-polarized (**d**) light.

## 1.4 Distinguishing The Signs of Doppler Shifts Using Polarization Degree of Freedom (DoF)

From the interference field (S13), one can see that for the x-polarized components, the interference terms refer to even cosine functions, whereas for the y-polarized components, related to odd sinusoidal functions. Thus, a fixed delay/advance in time may form in the Doppler time-domain spectra between these two polarized interference light. It gives the specific relative phase difference (RPDs) in the Doppler shift peaks in the Fourier spectra between these two polarized signals, namely, $\Delta\varphi_j = \varphi_y - \varphi_x = \tau\sigma\pi/2$ for $\Delta\omega_j > 0$, while $\Delta\varphi_j = -\tau\sigma\pi/2$ for $\Delta\omega_j < 0$, where $j$ = 1, 2, and 3. Note that $\Delta\varphi_4$ is always zero because this Doppler shift is determined by the rotational component alone. It originates from the direct amplitude (intensity) modulation when the moving particle interacts with the superposed field between the twisted components $\ell_1$ and $\ell_2$ of measuring structured light, but not from interference with the reference light. Hence, this Doppler shift $\Delta\omega_4$ is independent of the local polarization states. All these phase relations have been marked and shown (Fig. S5 and S6). The RDP $\Delta\varphi_4 = 0$, and others are $90°$ between x- and y-polarized interference light under the movement with $v = 1$ mm/s and $\Omega = 100 \times 2\pi$ rad/s (Fig. S5). By contrast, $\Delta\varphi_4 = 0$ and others are $-90°$ between these two polarized interference light when the movement is $v = -1$ mm/s and $\Omega = -100 \times 2\pi$ rad/s (Fig. S6). In return, this phenomenon can be used for distinguishing the signs of Doppler shifts $\Delta\omega_k$, $k = 1, 2, 3$.

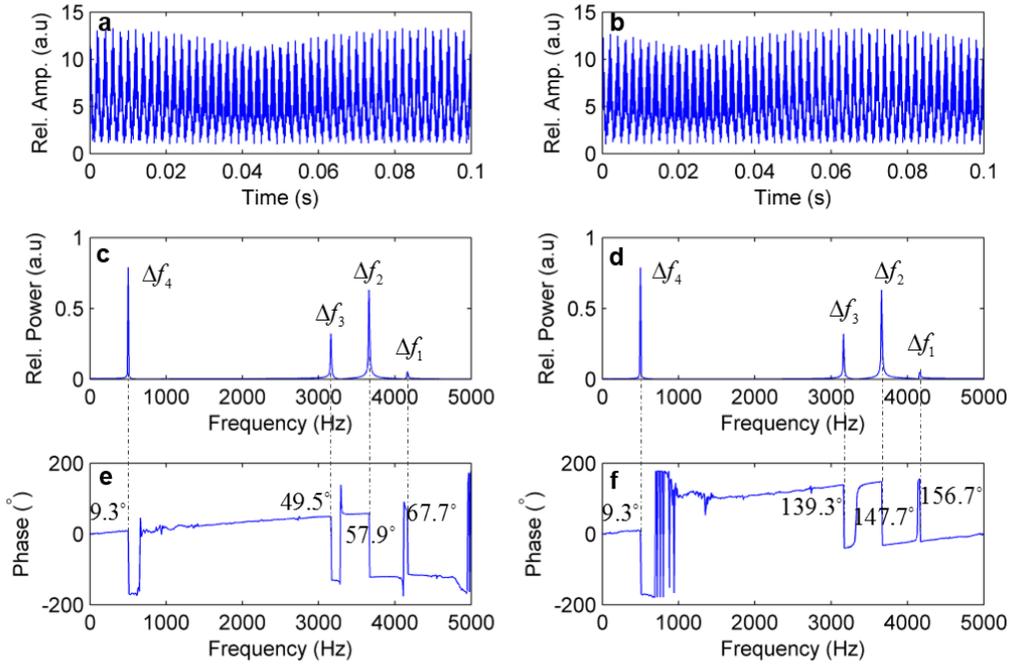

**Fig. S5 Calculated results for multi-dimensional movement with** $v = 1$ mm/s **and** $\Omega = 100 \times 2\pi$ rad/s **, under inclined angle** $\theta \simeq 0$**, rotation radius** $a = 1$ mm**, and** $\ell_1 = 5$**,**

$\ell_2 = 0$. **a, b** Doppler time-varying signals for received x-polarized (**a**) and y-polarized (**b**) light. **c, d** Fourier frequency spectra for received x-polarized (**c**) and y-polarized (**d**) light. **e, f** Fourier phase spectra for received x-polarized (**e**) and y-polarized (**f**) light, where the Doppler peak shifts between x-polarized (**e**) and y-polarized (**f**) light show fixed RPDs ($90°$), except for $\Delta\varphi_3$ that is zero.

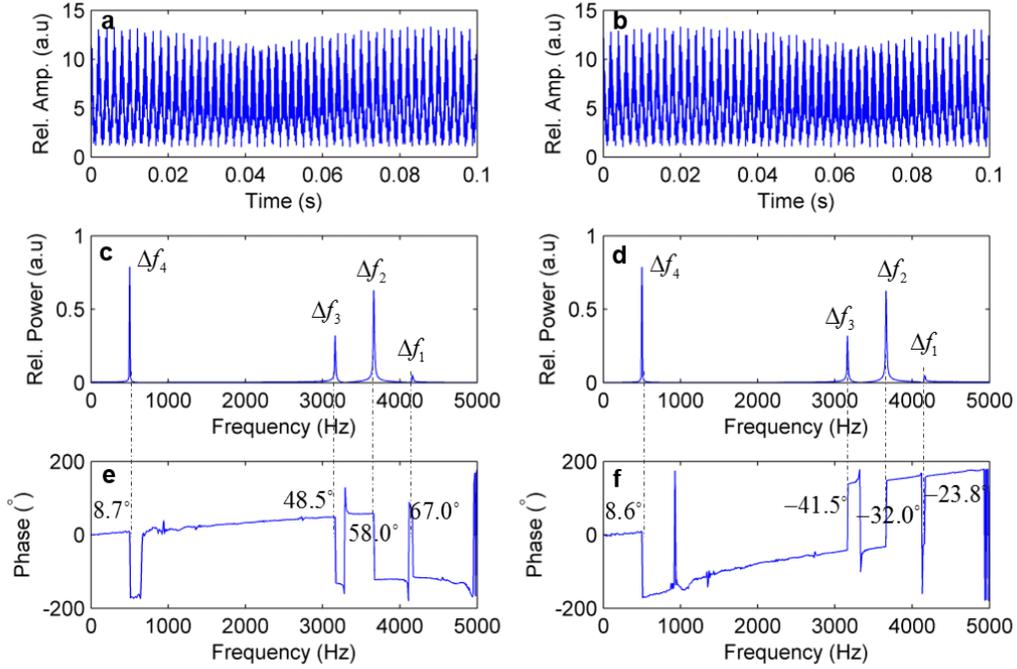

**Fig. S6 Calculated results for multi-dimensional movement with** $v = -1$ mm/s **and** $\Omega = -100 \times 2\pi$ rad/s **, under inclined angle** $\theta \simeq 0$**, rotation radius** $a = 1$ mm **, and** $\ell_1 = 5$**,** $\ell_2 = 0$. **a, b** Doppler time-varying signals for received x-polarized (**a**) and y-polarized (**b**) light. **c, d** Fourier frequency spectra for received x-polarized (**c**) and y-polarized (**d**) light. **e, f** Fourier phase spectra for received x-polarized (**e**) and y-polarized (**f**) light, where the Doppler peak shifts between x-polarized (**e**) and y-polarized (**f**) light show fixed RPDs ($-90°$), except for $\Delta\varphi_3$ that is zero.

### 1.5 Completely Experimental Setup for Multi-dimensional Measurement

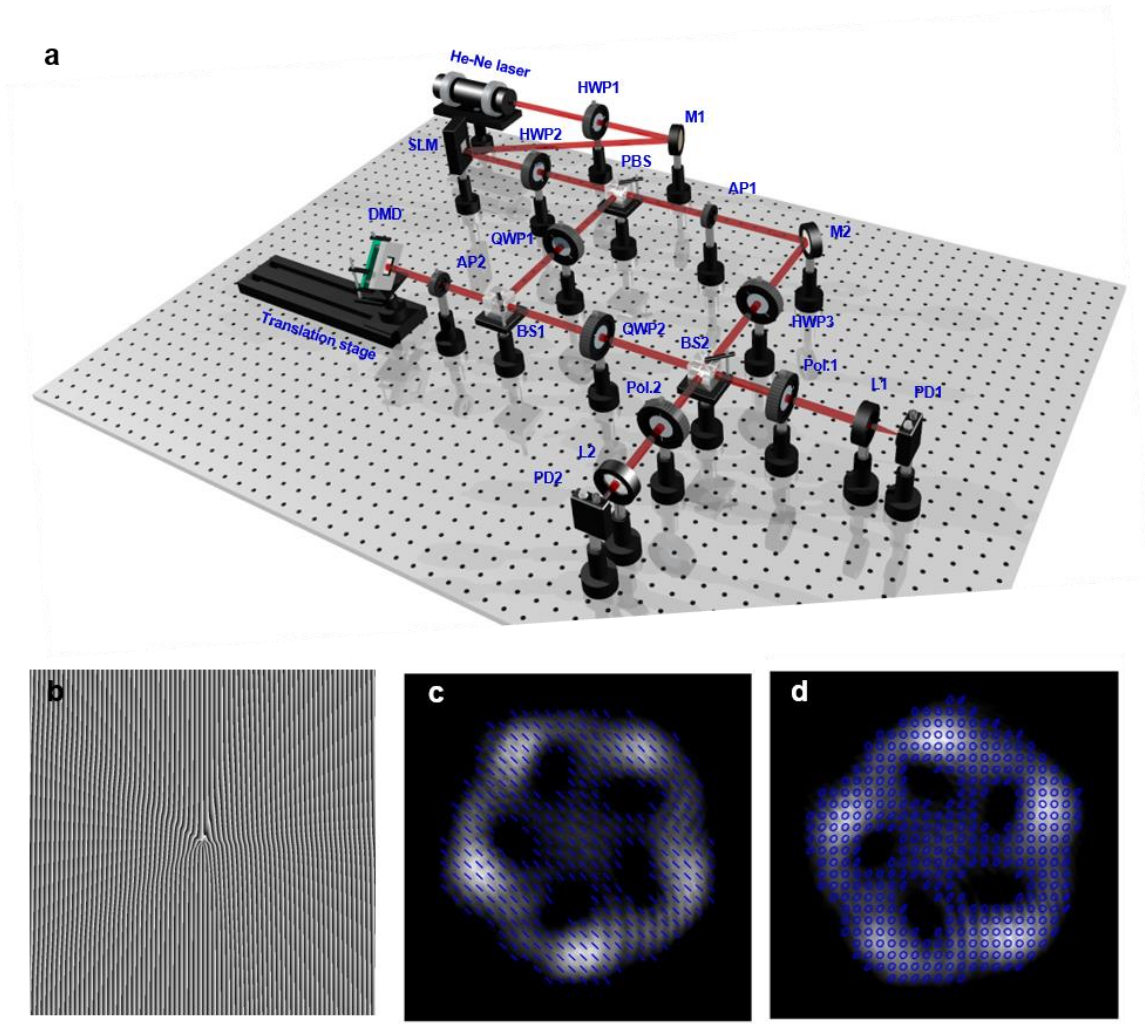

**Fig. S7 Experimental setup and generated structured light. a** Completely experimental setup. BS: beam splitter; PBS: polarization beam splitter; SLM: spatial light modulator; DMD: digital micromirror device; QWP: quarter wave plate; HWP: half wave plate; PD: photoelectric detector; L: lens; Pol.: polarizer; AP: aperture. **b** Forked complex phase pattern to generate the structured light ($\ell = 0, 5$). **c** Generated structured light with diagonal SoP as reference light. **d** Generated structured light with circular SoP as measuring light. The straight and circular blue lines represent the recovered diagonal and circular SoPs of structured light, respectively.

The complete experimental setup to detect the multi-dimensional movement of a moving particle is shown as (Fig. S7a). The structured light ($\ell = 0, 5$) was generated by the spatial light modulator (SLM) uploaded with forked complex phase pattern. In the experiment, the power allocation between two twisted components 0 and 5 was about 1:2.5. This structured light was split into two measuring and reference light, and the power radio of measuring to reference light was controlled about 20. The measuring structured light (Fig. S7c) was controlled to the circular SoP, while the reference light (Fig. S7b) was to the diagonal SoP. In the experiment, the moving particle was mimicked by a digital micromirror device (DMD). First, when reflected by the moving particle from the measuring structured light, a diffraction order was selected as the monitored light. This reflected light should be controlled to always parallel with the incident structured light by subtly adjusting the work plane of the DMD and the location of the sliding guide where the DMD was put, no matter how to translate the DMD on the sliding guide.

When conducting the measurement, the rotation center of the moving particle was controlled to nearly superpose with the axis of the measuring structured light by transversely moving the DMD. In the experiment, the rotation radius of moving particle was set about 1 mm, so the structured light was subtly scaled to match this rotation radius in order to get the amplitude-distinguishable Doppler shift peaks. Finally, the reflected light by the moving particle from the structured light was referred with the reference light, and then was split into two paths. The interference signals in two paths were filtered by two polarizers with orthogonally polarizing angles (x- and y-polarization), and then collected by two corresponding photoelectric detectors (PDs).

## 1.6 Additional Data for Measuring Multi-Dimensional Movement in different States

We present other measured results for multi-dimensional movement in another two different states (Fig. S8) in contrast with the measured results (Figs.2b-2g). Furthermore, we also give all the measured Doppler shifts and RPDs (Figs. S9 and S10) when retrieving the multi-dimensional velocities with four kinds of linear variation (Fig. 3a). The experimentally expectable variation of Doppler shift and difference with movement states fully demonstrate the feasibility of multi-dimensional measurement based on structured light by analyzing multiple Doppler shifts and RPDs.

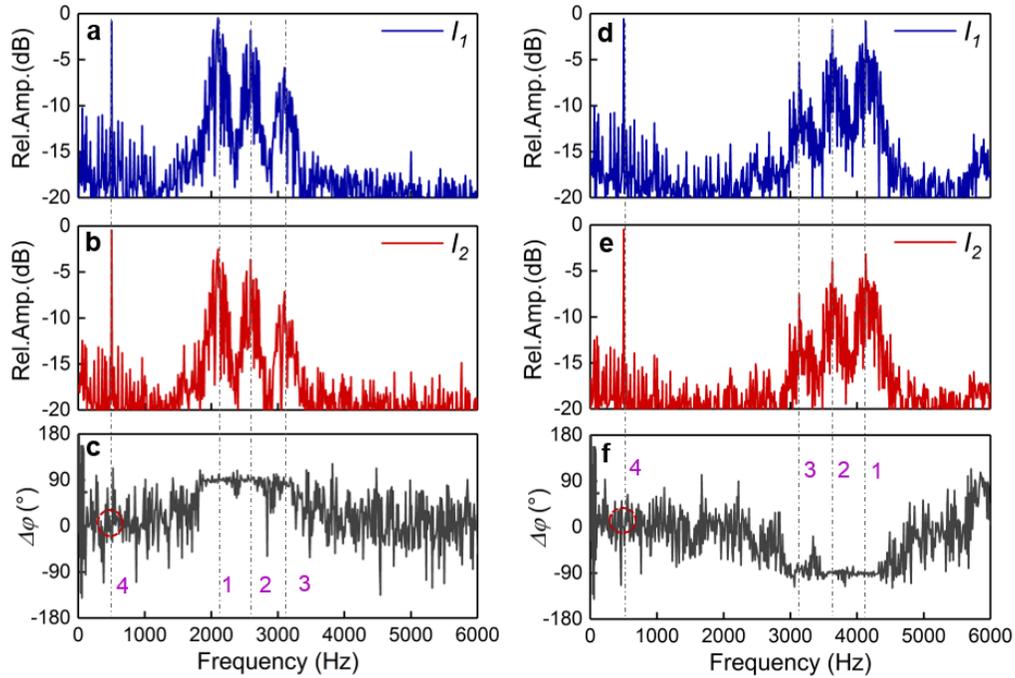

**Fig. S8 Measured Doppler frequency spectra for multi-dimensional movement with two different states. a-c** Measured results under $v_z = 1\,\mu\text{m/s}$ and $\Omega = -200\pi$ rad/s. **d-f** Measured results under $v_z = -1\,\mu\text{m/s}$ and $\Omega = -200\pi$ rad/s. **a, d** Doppler frequency-amplitude spectra ($I_1$) in the state of x-polarization by $P_1$ and $PD_1$. **b, e** Doppler frequency-amplitude spectra ($I_2$) in the state of y-polarization by $P_2$ and $PD_2$. **c, f** Doppler frequency-phase spectra as RPDs between these two Doppler signals with orthogonal SoPs (**a** and **b**, or **d** and **e**).

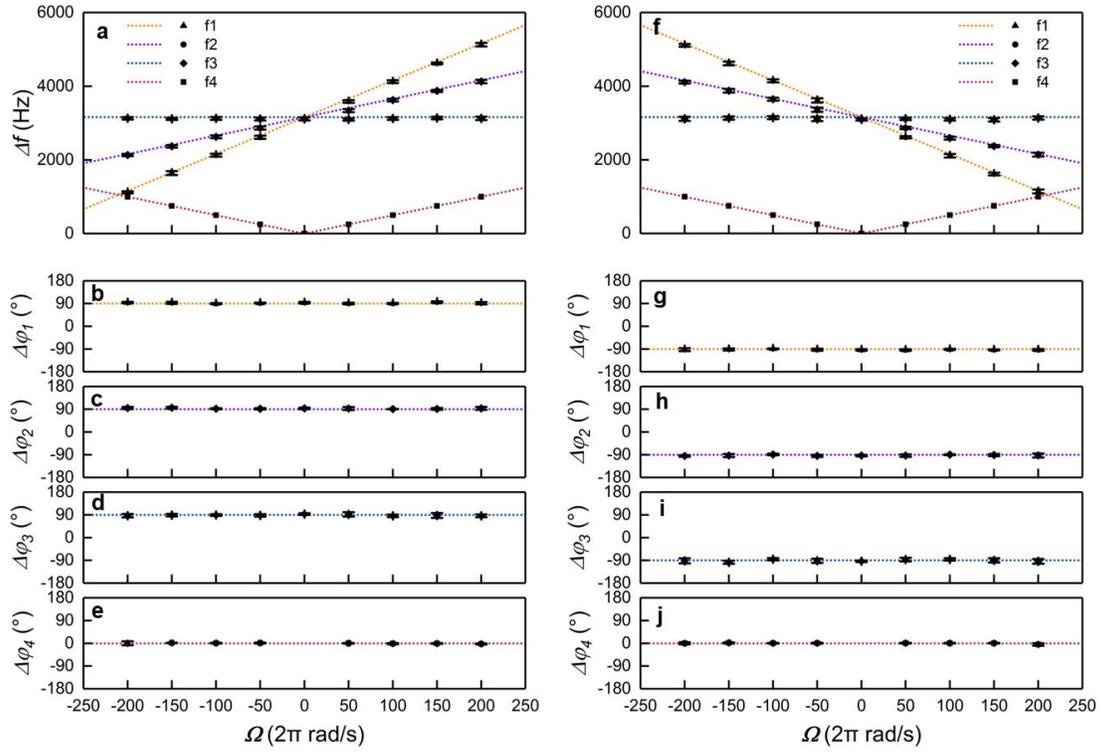

**Fig. S9 Measured Doppler shifts and RPDs versus different rotational velocities under fixed translational velocities. a** Measured Doppler shifts as experimental observables under fixed $v_z = 1\,\text{mm/s}$. **b-e** The corresponding RPDs between the orthogonally polarized Doppler signals. The experimentally retrieved multi-dimensional velocities. **f** Measured Doppler shifts as experimental observables under fixed $v_z = -1\,\text{mm/s}$. **h-j** The corresponding RPDs between the orthogonally polarized Doppler signals.

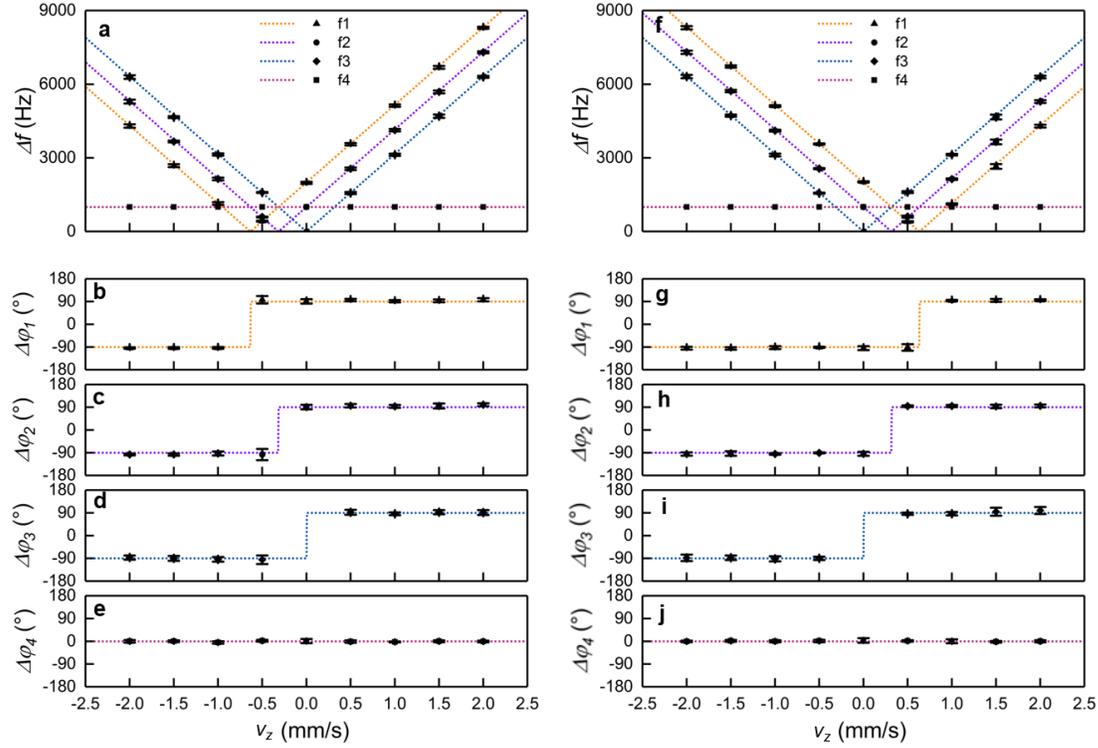

**Fig. S10 Measured Doppler shifts and RPDs versus different translational velocities under fixed rotational velocities. a** Measured Doppler shifts as experimental observables under fixed $\Omega = 400\pi$ rad/s . **b-e** The corresponding RPDs between the orthogonally polarized Doppler signals. The experimentally retrieved multi-dimensional velocities. **f** Measured Doppler shifts as experimental observables under fixed $\Omega = -400\pi$ rad/s . **h-j** The corresponding RPDs between the orthogonally polarized Doppler signals.

## Derivation of Doppler effect Induced by Moving Bulk Targets

### 2.1 Doppler Shift of Single Twisted Light by Moving phase Masks

As for the incident plane-phase light propagating along the $+z$ direction, its electric-field function can be simply given by

$$E(z,t) = A_0(r)\exp\left[i(\omega t - kz)\right]$$

(S19)

where $A_0(r)$ is the complex amplitude of transverse electric-field, $r$ is the radial position away from the axis of light. A helical phase mask following with a moving target can be expressed as a transmission function of $H(\theta) = \exp(i\ell\theta)$. As for a reflection-type mask, the illuminating plane-phase light can be modulated into the twisted light,

$$E_\ell(\theta) = E(-z,t)\exp(i\ell\theta). \tag{S20}$$

If this mask following with the moving target is in the state of multidimensional movement, including translational and rotational components, considering that both the translation direction and the rotation axis direct along/against the axis of light, its transmission function can be modified as

$$H(\Omega) = \exp\left[i\ell(\theta+\Omega t)\right]\exp(i2kv_z t). \tag{S21}$$

Accordingly, the twisted light reflecting off this moving mask can be written as

$$E_\ell(\theta,v,\Omega) = E \cdot H = A_0(r)\exp(i\ell\theta)\exp\left[i(\omega+\ell\Omega+2kv_z)t\right]\exp(ikz). \tag{S22}$$

The coaxial interference between the twisted light $E_\ell(\theta,v,\Omega)$ and a reference light $E_{ref} = A'(r)\exp(i\omega t)\exp(ikz)$ can be approximately given as

$$I = \left(E_\ell + E_{ref}\right)^* \cdot \left(E_\ell + E_{ref}\right) = |A_0|^2 + |A'|^2 + 2A_0 A'\cos\left\{\ell\left[\theta + \frac{1}{\ell}(\ell\Omega + 2kv_z)t\right]\right\}, \tag{S23}$$

which gives a helical interference pattern with the number of helical fringes of $|\ell|$. From Eq. S23, both the translational ($v_z$) and rotational ($\Omega$) components of the moving phase mask can induce the interference pattern rotating. The rotational velocity ($\Omega'$) of each helical fringe is

$$\Omega' = \frac{\ell\,\Omega + 2kv_z}{\ell}. \tag{S24}$$

When monitoring a fixed off-axis position in this interference pattern, a periodically time-varying signal can be gotten, giving the overall Doppler shift,

$$\Delta\omega = \ell\Omega' = \ell\,\Omega + 2kv_z, \tag{S25}$$

which contains both translational and rotational Doppler contributions. It is worth noting that the rotational and translational components cannot be distinguished and determined by analyzing this Doppler shift, because of their equivalent contributions.

### 2.2  Doppler shifts of General Structured light by Moving Masks (Targets)

In general, the structured light as the superposition of multiple twisted components ($\ell_1$, $\ell_2$, ..., $\ell_n$) can be generated by a complex phase mask. Firstly, we present the feasible methods to design the complex phase-only masks to generate the superposing twisted light with high purity by suppressing undesired twisted components. Considering the reflective type of phase masks, the targeted transmission function of them can be expressed as

$$H(\theta) = \sum_{m=1}^{n} A_m \exp(i\ell_m\theta). \tag{S26}$$

The approximate phase-only function can be defined as $H(\theta) = \exp[i\varphi(\theta)]$ with a phase function [S1],

$$\varphi(\theta) = \mathrm{Re}\left\{-i\ln\left[\sum_{m=1}^{n} B_m \exp(i\ell_m\theta)\right]\right\}, \tag{S27}$$

where Re{ } means "real part of", and $B_m$ is the decisive factor of each helical phase component. $H(\theta)$ can be written as Fourier series,

$$H(\theta) = \sum_{m=-\infty}^{\infty} F_m \exp(i\ell_m \theta), \tag{S28}$$

where $F_m = \int_0^{2\pi} H(\theta) \exp(-im\theta) d\theta / 2\pi$ is the decomposition coefficient. One can use an iterative algorithm to obtain the optimized $F_m$ by minimizing the relative root-mean-square error ($R\text{-}RMSE$)

$$R\text{-}RMSE = \sqrt{\frac{\sum_{m=1}^{n}\left(|F_m|^2 - |A_m|^2\right)^2}{n\sum_{m=1}^{n}|F_m|^2}}. \tag{S29}$$

As for the superposition of two twisted components ($\ell_1$, $\ell_2$) that is employed to the structured light interferometry, we assume that this transmission function of complex phase mask has been optimized as $H(\theta) = \sum_{m=1,2} F_m \exp(i\ell_m \theta)$. When this complex mask attached on a moving target is in the state of both translational velocity $v_z$ and rotational velocity $\Omega$, this optimized transmission function can be given by

$$H(\theta, v, \Omega) = \sum_{m=1,2} F_m \exp\left[i\ell_m(\theta + \Omega t)\right] \exp(i2kv_z t). \tag{S29}$$

Accordingly, the superposed field generated by the moving mask can be expressed as

$$E_{\ell_1, \ell_2} = E(r,-z,t) \cdot H(\theta, v_z, \Omega) = \sum_{m=1,2} F_m A_0 \exp(i\ell_m \theta) \exp\left[i(2\pi f_0 + \ell_m \Omega + 2kv_z)t\right] \exp(ikz) \tag{S30}$$

This superposed field with equal weight coefficients ($F_1 = F_2$) produces a multi-petal interference pattern, of which the intensity distribution can be approximately written as

$$I_{\ell_1,\ell_2} = \left|E_{\ell_1,\ell_2}\right|^2 \propto 1 + \cos\left[\Delta\ell(\theta + \Omega t)\right], \tag{S31}$$

where $\Delta\ell=\ell_2-\ell_1$, and its absolute $|\Delta\ell|=|\ell_2-\ell_1|$ directs the number of petals. From Eq. S31, just the rotational component ($\Omega$) of the complex phase mask directly gives the multi-petal interference pattern rotation with the velocity $\Omega'=\Omega$. When monitoring a fixed off-axis position in this interference pattern, similarly, a periodic time-varying intensity can be observed as the well-known rotational Doppler shift,

$$\Delta\omega=|\Delta\ell|\Omega. \tag{S32}$$

In general, these two twisted components may have different winding numbers, $|\ell_1|\neq|\ell_2|$. When coaxially interfering with a reference plane-phase light, a composite interference pattern can be formed with three kinds of interference regions, i.e. the inner region I with interference between the twisted component $\ell_1$ and the reference plane-phase light, the middle region II as the superposition of twisted components $\ell_1$ and $\ell_2$, and the outer region III with interference between the twisted component $\ell_2$ and the reference plane-phase light. The region I and region III feature helical fringes and the region II shows multi-petal fringes. According to the analysis above, three Doppler shifts can be gotten when off-axis monitoring the composite interference pattern. The regions I, II, and III give $\Delta\omega_1=\ell_1\Omega+2kv_z$, $\Delta\omega_2=|\ell_2-\ell_1|\Omega$, and $\Delta\omega_3=\ell_2\Omega+2kv_z$, respectively. In practice, two of them are enough to determine $v_z$ and $\Omega$ by solving the linear equations in two unknowns. Therefore, in the experiment, the outermost interference region was removed by suitably shrinking the size of reference plane-phase light to just get the composite interference pattern with two regions I and II.

### 2.3 Completely Experimental Setup for Multi-dimensional Measurement

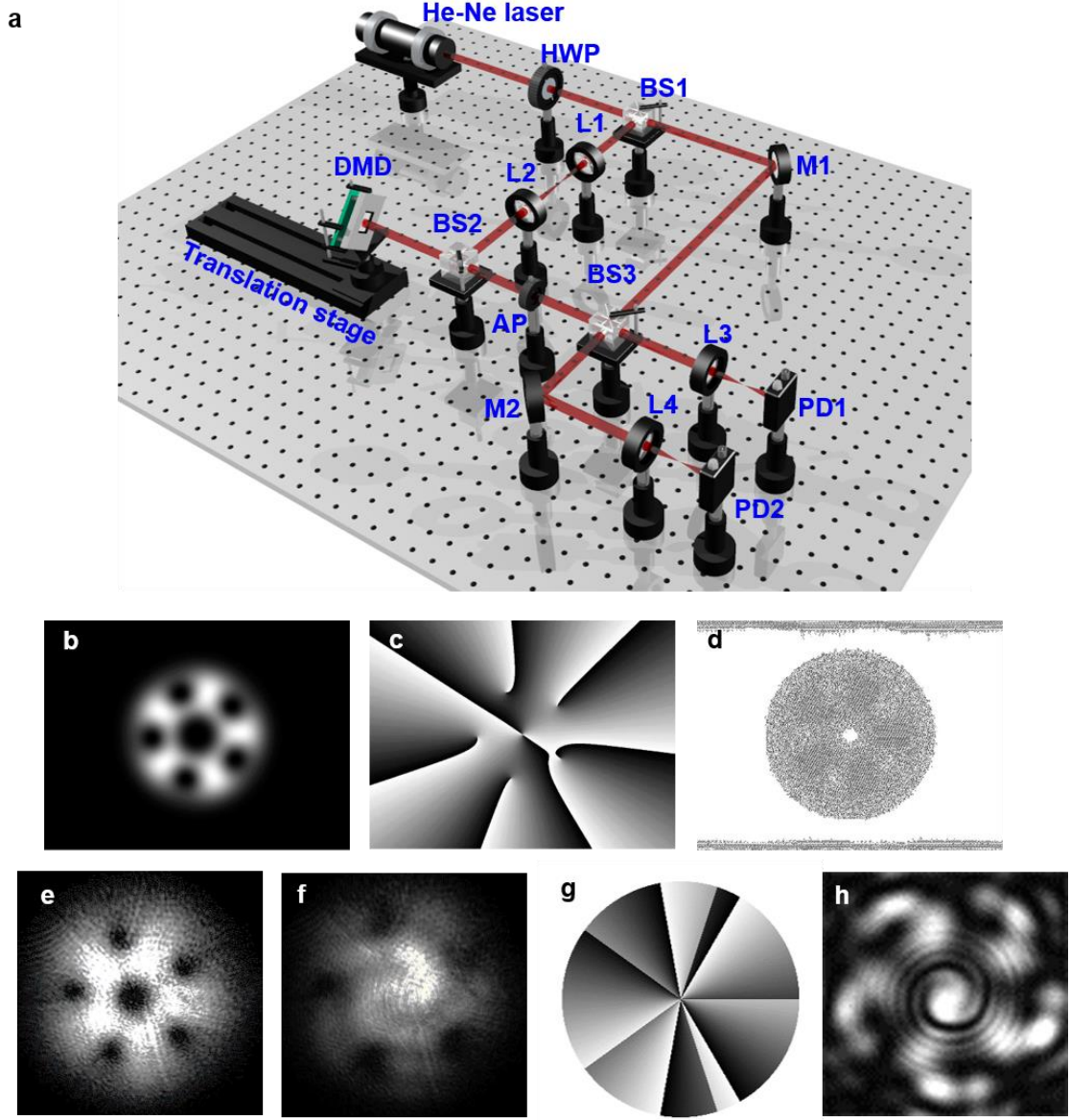

**Fig. S11 Experimental setup and generated structured light. a** Completely experimental setup. BS: beam splitter; DMD: digital micromirror device; HWP: half wave plate; PD: photoelectric detector; L: lens; AP: aperture. **b**, **c** Calculated intensity and phase ($\ell = 2, 7$) using the superpixel method prepared for getting DMD holographic pattern. **d** The resulting DMD holographic pattern to generate structured light ($\ell = 2, 7$). **e** Experimentally generated structured light. **f** The interference pattern between the generated structured light and the reference Gaussian beam. **g** The true complex phase mask installed on the moving target is equivalent to the DMD holographic pattern (**d**) for the practically multi-dimensional movement.

**h** The interference pattern between the structured light ($\ell = 2, 7$) generated by the complex phase mask (**g**) and reference Gaussian beam, which is equivalent to the interference pattern (**f**) when using for true measurement

The complete experimental setup is shown to simulate the multi-dimensional measurement of a moving bulk target (Fig. S11a). The Gaussian beam was split into two measuring and reference light. The measuring Gaussian beam was expended about 10 times through $L_1$ and $L_2$ before illuminating on the DMD. Here we took the superpixel methods to generate structured light ($\ell = 2, 7$) using the DMD [S2]. The calculated intensity and phase were beforehand prepared to get the resulting DMD holographic pattern (Fig. S11b-S11d). The axis of measuring Gaussian beam should be kept to nearly superpose with the center of the DMD holographic pattern to generate the high-quality structured light. After one diffraction order of generated structured light was selected as the monitored structured light (Fig. S11e), this diffraction order should be controlled to always parallel with the incident structured light by subtly adjusting the work plane of the DMD and the location of the sliding guide where the DMD was put, no matter how to translate the DMD on the sliding guide. Finally, the monitored structured light was referred with the reference Gaussian beam to get the interference pattern (Fig. S11f). This interference pattern was split two paths, and then expended so that the screens of two PDs just can collect the local light in the structured interference field. These two DPs were positioned at the junction of two interference regions with a relative azimuthal position to collect the Doppler signals. In this scheme, for experimental convenience, we replaced the true complex phase mask (g) installed on the moving target by the equivalent DMD holographic pattern (d). In the true measurement, the interference pattern (f) would be replaced by true interference pattern (h) used for extracting Doppler signals.

## 2.4 Additional Results for Different Kinds of Multi-Dimensional Motion

We present other measured results for multi-dimensional movement in another two different states (Fig. S12) in contrast with the measured results (Figs.4b-4g). Furthermore, we also give all the measured Doppler shifts and RPDs (Figs. S13 and S14) when retrieving the multi-dimensional velocities with four kinds of linear variation (Fig. 5a). The experimentally expectable variation of Doppler shift and difference with movement states fully demonstrate the feasibility of multi-dimensional measurement based on structured light by analyzing multiple Doppler shifts and RPDs.

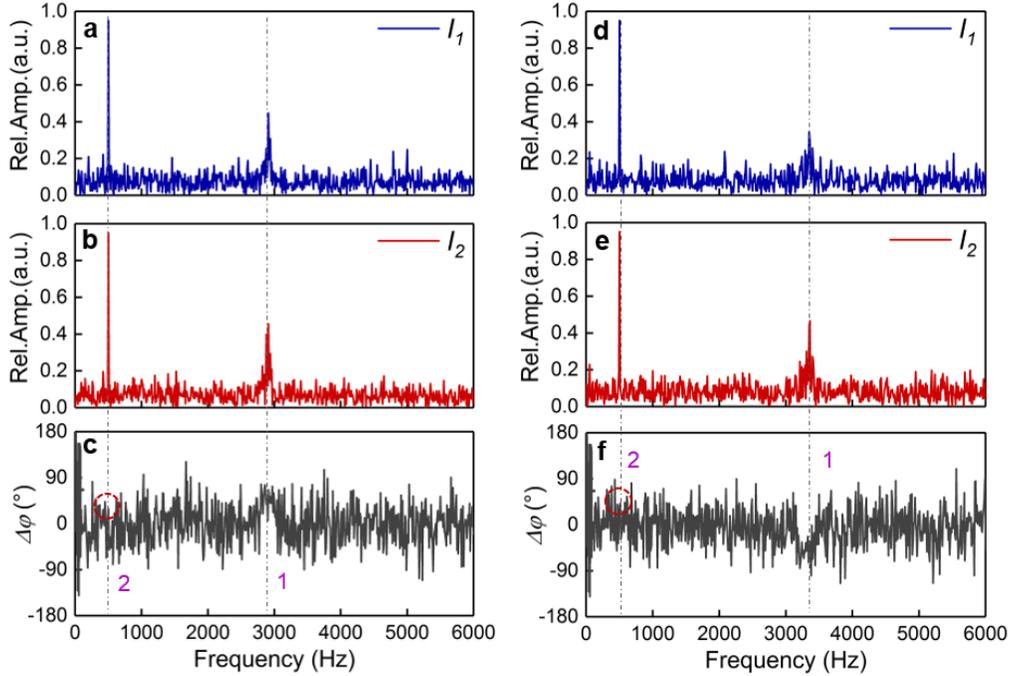

**Fig. S12 Measured Doppler frequency spectra for multi-dimensional movement with two different states. a-c** Measured results under $v_z = 1\,\mu\text{m/s}$ and $\Omega = -200\pi$ rad/s. **d-f** Measured results under $v_z = -1\,\mu\text{m/s}$ and $\Omega = -200\pi$ rad/s. **a, d** Doppler frequency-

amplitude spectra (I$_1$) gotten from L$_1$ and PD$_1$. **b, e** Doppler frequency-amplitude spectra (I$_2$) gotten from L$_2$ and PD$_2$. **c, f** Doppler frequency-phase spectra as RPDs between these two Doppler signals at two relative azimuthal positions (**a** and **b**, or **d** and **e**).

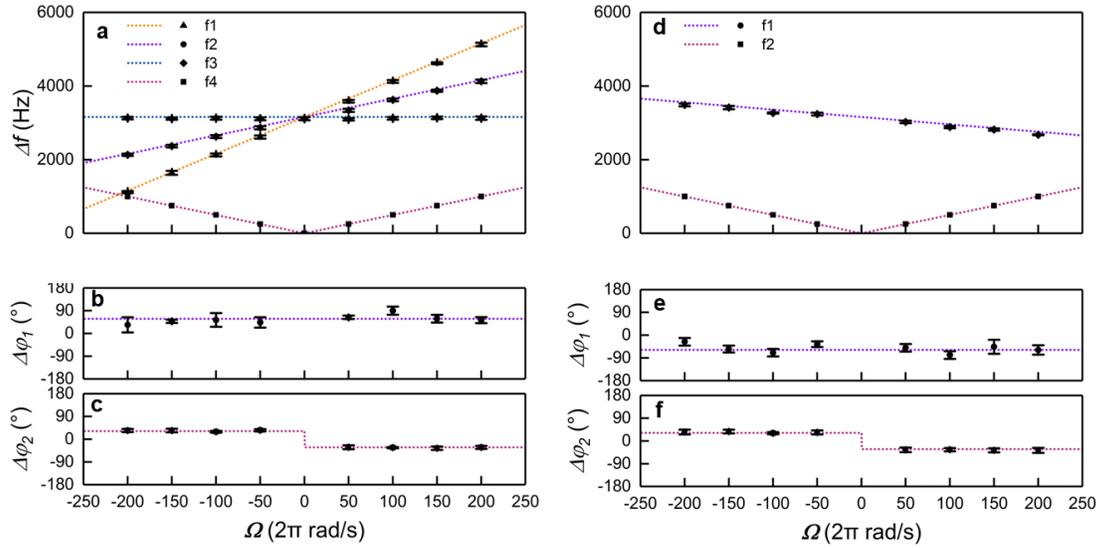

**Fig. S13 Measured Doppler shifts and RPDs versus different rotational velocities under fixed translational velocities. a** Measured Doppler shifts as experimental observables under fixed $v_z = 1\,\text{mm/s}$. **b, c** The corresponding RPDs between Doppler signals at two relative azimuthal positions. The experimentally retrieved multi-dimensional velocities. **d** Measured Doppler shifts as experimental observables under fixed $v_z = -1\,\text{mm/s}$. **e, f** The corresponding RPDs between between Doppler signals at two relative azimuthal positions.

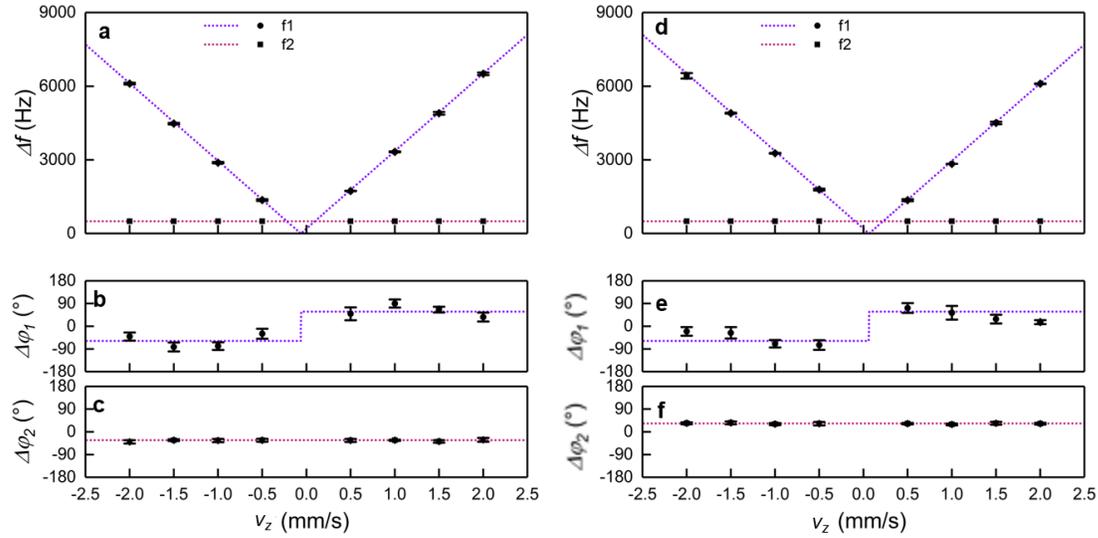

**Fig. S14 Measured Doppler shifts and RPDs versus different translational velocities under fixed rotational velocities. a** Measured Doppler shifts as experimental observables under fixed $\Omega = 200\pi$ rad/s. **b, c** The corresponding RPDs between Doppler signals at two relative azimuthal positions. The experimentally retrieved multi-dimensional velocities. **d** Measured Doppler shifts as experimental observables under fixed $\Omega = -200\pi$ rad/s. **e, f** The corresponding RPDs between Doppler signals at two relative azimuthal positions.

**Restriction Condition of Multi-dimensional Measurement**

In both interferometric schemes based on structured light to detect a moving objects, including both small particles and bulk objects, in principle, the detectably multi-dimensional movement should be conditioned with the translational direction of the moving objects being nearly collinear with the rotation axis of them. Otherwise for the case of detecting the moving particles, the deduced Doppler shifts would be no longer constants in spite of the fixed translational and rotational velocities, but become time-varying, due to the motion perturbation resulted from the inclined rotation plane (see the analysis and discussion in the first part). In order to get the distinguishable Doppler shifts, the tolerable inclined angle $\theta$ between the translational direction and the rotation axis is less than $\lambda/4a$, where $a$ is the rotation radius of the moving particle.

Nevertheless, in the experiment, as for detecting the moving particle, that collinear condition above actually was not satisfied. Because the rotational plane of the mimicked particles must be an inclined angle $7°$ with respect to the translational direction for the measuring structured light that can be reflected back still along the axis of light by the moving particle. This is inherently restricted by the DMD where the micromirrors are in the state of 'ON', giving the inclined angle $7°$. However, in the experiment, the measurement had not been affected by this inclined angle. This is because the moving particle was mimicked by DMD through changing the time interval when continuously switching the next set of micromirrors to the 'ON' state. In this case, the longitudinal movement component of the mimicked particle resulted from the inclined rotation plane was not really in the state of continuous motion as the true particle, which did not cause the longitudinal Doppler shift perturbation.

As for the case of measuring the moving bulk targets, the central normal of the mask surface should be kept collinear with the rotation axis when installing it on the bulk target, and the measuring Gaussian beam should be illuminated on the center of complex phase mask. Otherwise, the structured light generated would be always shifted so that cannot be interfered with reference light, or the interference pattern cannot form interference pattern with two clear regions to distinguish the signs of Doppler shifts. In the experiment, the complex phase mask installed on moving targets was replaced by uploading DMD holographic patterns. The inclined plane of holographic patterns due to the inherent inclination of micromirrors on the DMD need not be considered provided that the reflection direction of the diffraction order monitored as the targeted structured light is adjusted to always parallel with the translational direction of DMD on this sliding guide. Actually, in the practical measurement, the designed holographic pattern can be directly used for measuring the moving bulk target when installing it on the target, which is equivalent to the case of using complex phase mask. The difference is the method of generating structured light, the holographic pattern generates structured light through both amplitude and phase modulation, while the complex phase mask does it through pure phase modulation. This difference does not affect the utilization of them to deliver motion information of the moving target so as to get the multi-dimensional measurement.

**Data Processing of Multi-dimensional Measurement**

For each state of multi-dimensional movement, we conducted multiple measurements (more than 5 times). The repeated data, on the one hand, were used for making tolerance analysis to get the error bar, on the other hand, significantly could be available to exclude noise frequencies and thus better identify the Doppler shift peaks in the Doppler frequency spectra. For the latter, first, we averaged all the multiply measured data of Doppler frequency-phase signals. The frequency ranges of targeted Doppler shifts could be roughly locked by judging the averaged RPDs at a certain range. Because for every measurement of one movement state, the targeted Doppler shifts almost give the fixed RPDs. However, other frequencies do not have the fixed RPDs, because these frequencies corresponding to noise frequencies are random for every measurement. When averaging these noise frequencies, their RPDs may approach zero. Therefore, based on this judgment method, we could effectively exclude the noise frequencies so as to clearly recognize the targeted Doppler shift peaks. This operation is useful in practical measurement. In this experiment, it is reasonable to identify this multiple measurements as continuously multiple data collection in the process of one measurement.

**Accuracy and Tolerance Analyses of Multi-dimensional Measurement**

In the experiment, the time window of collecting the measured data was 0~0.1 seconds, and the sampling amount was 2500 when collecting the measured data. This gave the frequency resolution of 10Hz, and the frequency window of 25000Hz. The frequency range (0~6000Hz) was displayed in the main text. This frequency resolution (10Hz) may give rise to about $\pm 5\lambda \text{ s}^{-1}$ of error range when retrieving the translational velocities. For example, when $\lambda = 632.8$ nm, the error range is about $\pm 3.2 \times 10^{-3}$ mm/s.

From the measured results, the Doppler shifts ($\Delta f_4$ in Case one and $\Delta f_2$ in Case two) show more accurate values than other Doppler shifts. Because this Doppler shift component could be directly extracted without by interference with reference light, and thus nearly not affected by ambient disturbance. It derives from the direct interaction of the rotational movement of objects with the superposed field between different twisted components of structured light, giving the rotational-only Doppler shift. So when retrieving the multi-dimensional velocities for Case one, the retrieved rotational velocities could be rectified through this Doppler shift ($\Delta f_4$). By contrast, other Doppler shift components in both cases have to be extracted by interference with reference light, and thus inevitably affected by the ambient disturbance. The resulting measured values show errors in a small range (about $\pm 150$ Hz), giving rise to the retrieved translational velocities with minor errors (about $\pm 0.05$ mm/s).

As for distinguishing Doppler shifts through the RPDs, the tolerance range of judgement is $180°$ if setting theoretical decision values $\Delta\varphi = \pm 90°$. For example, if the measured RPDs fall into the range from 0 to $180°$ (or from $-180°$ to 0), one can give the judgement that the measured Doppler shifts have plus (or minus) signs. For the case of detecting small particles by exploiting the polarization DoF of structured light, from the experimental results, all the measured RPDs gave nearly the exact values $\pm 90°$. Because except for the QWPs and HWPs to control the SoP, other devices without birefringence used in the experimental configuration did not affect the SoPs of the orthogonal polarized Doppler signals. For the case of measuring bulk targets, the measured $\Delta\varphi_1$ around $\pm 70°$ had large errors with the prediction, which results from the ambient disturbance on inner interference fringes, as well as the rough relative positions when putting two PDs. Nevertheless, these errors had not influence on the distinguishment of the signs of Doppler shifts because of the large tolerance range.

## Comparison among Four Optical Interferometric Schemes Using different kinds of light

The different interferometric schemes discussed here are suitable for both metrology scenarios of small particle and bulk target. Here we just take the scenarios of measuring the bulk target as the example to compare the functionality among these schemes using different kinds of light.

### 1.1 Plane-Phase Light

In the conventional interferometry, the mirror can be identified as a constant phase mask, the plane-phase (fundamental Gaussian) light reflected by the mirror following with moving objects gives the detectable linear Doppler shift, i.e. $\Delta\omega=2kv_z$. When coaxially interfering this light with the reference plane-phase light, a concentric ring interferogram or Newton's rings can be formed for the measurement of only translational movement (Fig. S15a). However, this conventional scheme fails to measure the rotational movement. Note that although both incident and reflected light need to be normal with respect to the plane of mirror, an angle between reflected and incident light is drawn for simplicity and clarity of conceptual illustration (Fig. S15a-15d).

### 1.2 Single Twisted Light

When the mirror is replaced by a helical phase mask (Fig. S15b), the incident plane-phase light can be spatially modulated to the twisted light. The interference between the twisted light and the reference plane-phase light features a helical interference pattern. In this case, both translational and rotational movement may cause the interference pattern rotation around its center. When monitoring a fixed off-axis position at the interference pattern, the rotating helical fringes can give the overall Doppler shift $\Delta\omega = \ell\,\Omega + 2kv_z$. This Doppler shift is contributed by both translational and rotational Doppler effect. Therefore, the helical phase mask supports independent measurement of translational movement under $\Omega=0$ or rotational movement under $v_z=0$. However, the determination of both translational and rotational components are pending from the overall Doppler shift.

### 1.3 Superposed Twisted Light with Opposite Topological Charges

If a complex phase mask with two opposite helical components (Fig. S15c), the incident plane-phase light can be converted into the superposition of two twisted light ($\ell$, $-\ell$) characterized by a petal-like interference pattern with the number of $|\Delta\ell|=2|\ell|$ petals. The rotational velocity of each petal-like interference fringe is the same as the rotating objects. When monitoring a fixed off-axis position at this petal-like interference pattern, the rotating fringe gives the well-known rotational Doppler shift written by $\Delta\omega = 2|\ell|\Omega$. This rotational Doppler shift depends only upon the rotational velocity of moving objects regardless of the translational movement.

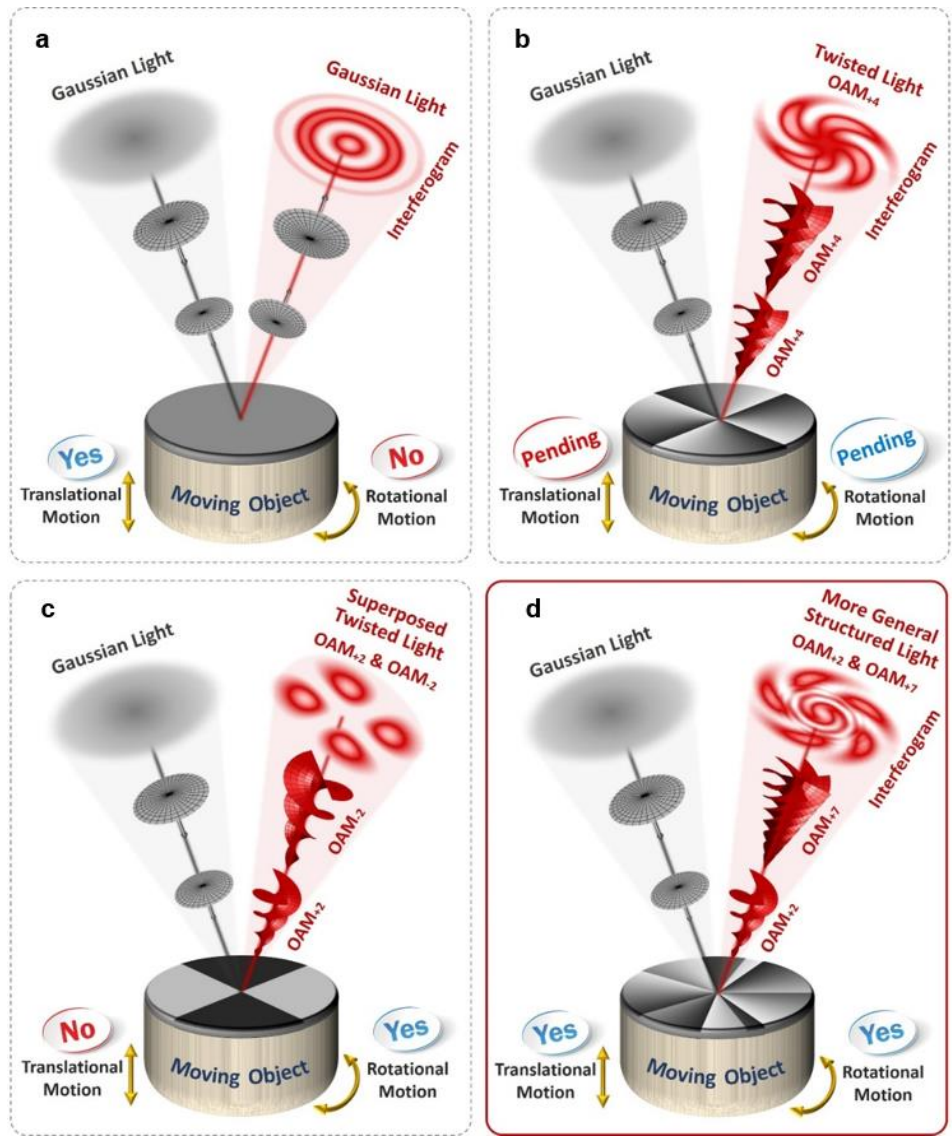

**Fig. S15. Conceptual illustration of four interferometric schemes using different kinds of light.** When conducting the measurement, a phase mask should be attached on the moving object to detect the movement information. **a** A mirror identified as a constant phase mask only allows the translational movement measurement but not the rotational movement. **b** A helical

phase mask generating single twisted light allows the independent measurement of only translational or rotational movement. **c** A complex phase mask generating superposed twisted light with opposite topological charges ( $\ell$ , $-\ell$ ) only allows the rotational movement measurement but not the translational movement. **d** More general complex phase mask generating superposed twisted light with different topological charges ( $|\ell_1| \neq |\ell_2|$ ). The interference pattern with a reference Gaussian light, forming a composite interference pattern, allows the full information extraction of multidimensional movement (translational velocity and rotational velocity, as well as the directions of them).

### 1.4 General Structured Light with Two Twisted Components ($|\ell_1| \neq |\ell_2|$)

This structured light interferometry using general structured light with two twisted components (Fig. S15d) has been theoretically and experimentally demonstrated in detail in the main text for both metrology scenarios of measuring small particles and bulk targets.

It should be stressed that for all schemes above, extracting Doppler signals by directly interference with reference light, one can only get the magnitude of the translational and/or rotational velocity, because the signs of Doppler shifts cannot be distinguished. The direction detection can be achieved by distinguishing the signs of Doppler shifts through the RPDs between two Doppler signal components in different polarized states or monitored in different positions at the interference patterns (see the main text).

### The Case of Conceptually Detecting a Moving Rough Surface

A rough surface can be described as a reflective transfer function $H(r,\theta)$ in polar coordinates. It can be expanded by a series of helical phase components $H(r,\theta) = \sum_m A_m(r) \exp(im\theta)$, where $A_m(r)$ is the modulation coefficient of each component, and can be normalized with $\sum_m \left| \int A_m(r) \, dr \right|^2 = 1$. When the rough surface moves with a rotational velocity $\Omega$ and a linear velocity $v_z$ (the direction of the linear motion is parallel to the normal of the rough surface), the modified phase function of the moving rough surface relative to the incident light can be written by

$$H(v, \Omega) = \sum_m A_m(r) \exp\left[ im(\theta + \Omega t) \right] \exp\left\{ i (\cos\alpha + \cos\beta) kvt \right\}$$
$$= \sum_m A_m(r) \exp\left[ im(\theta + \Omega t) \right] \exp\left\{ i \left[ 2 - \frac{1}{2}(\alpha^2 + \beta^2) \right] kvt \right\},$$

(33)

where $\alpha$ and $\beta$ indicate the small incident angle and small scattered angle (detection angle) with respect to the normal of the rough surface, respectively. Note that here we make an approximation $\cos\alpha = 1 - 2\sin^2(\alpha/2) \approx 1 - \alpha^2/2$, and similarly for $\beta$.

The electric-field function of an incident light beam with frequency $f$ can be given by

$$E_1 = A_1(r) \exp\left[ i(2\pi f t) \right] \mathbf{e}_1,$$

(34)

where $\mathbf{e}_1 = \begin{bmatrix} 1 & 0 \end{bmatrix}^T$ denotes x-polarization, $A_1(r)$ is related to the transverse electric field distribution of the light beam and the relevant phase term of light propagation is not given here. The scattered light from the moving rough surface can be expressed as

$$E_2(m) = E_1 \cdot H = \sum_m A_m A_1 \exp(im\theta) \exp\left\{ i \left[ 2\pi f + m\Omega + 2kv - \frac{1}{2}(\alpha^2 + \beta^2) kv \right] t \right\} \mathbf{e}_1,$$

(35)

which shows that different twisted components give different Doppler shifts. If a specific helical mask ($-\ell$) was employed to convert these twisted components and then the converted plane-phase light ($\ell \to 0$) was collected by a single-mode fiber (SMF), this collected light component could be used for extracting the Doppler shift (including magnitude and direction) induced by the moving rough surface (Fig. S16). The subsequent orthogonal polarization detection to get the sign-distinguishable Doppler shift is similar to the Case one when detecting the moving small particle. The whole process from converting scattered light component to get the circularly polarized Doppler signal can be formulized as,

$$E_3 = \begin{bmatrix} 1 & -i \\ -i & 1 \end{bmatrix} E_2(m=\ell) \cdot \exp(-i\ell\theta) = \\ = \begin{bmatrix} 1 \\ -i \end{bmatrix} A_\ell A_1 \exp\left\{i\left[2\pi f + m\Omega + 2kv - \frac{1}{2}(\alpha^2+\beta^2)kv\right]t\right\}. \tag{36}$$

If referred with a reference light with diagonal polarization,

$$E_0 = A_0 \exp\left[i(2\pi ft)\right]\mathbf{e}_0, \tag{37}$$

where $\mathbf{e}_0 = \begin{bmatrix} 1 & 1 \end{bmatrix}^\mathbf{T}$, and then the polarized Doppler interference signal is split into two paths where two polarizers with x- and y polarizing directions are positioned, one can get two orthogonally polarized Doppler signals, respectively given by

$$I_{\ell,x} = \left\|\begin{bmatrix} 1 & 0 \\ 0 & 0 \end{bmatrix} \cdot (E_0 + E_3)\right\|^2 = A_0^2 + A_1^2 A_\ell^2 + 2A_0 A_1 A_\ell \cos\left[\left(\ell\Omega + 2kv - \frac{1}{2}(\alpha^2+\beta^2)kv\right)t\right],$$

(38)

and

$$I_{\ell,y} = \left\|\begin{bmatrix} 0 & 0 \\ 0 & 1 \end{bmatrix} \cdot (E_0 + E_3)\right\|^2 = A_0^2 + A_1^2 A_\ell^2 + 2A_0 A_1 A_\ell \cos\left[\left(\ell\Omega + 2kv - \frac{1}{2}(\alpha^2+\beta^2)kv\right)t + \frac{\pi}{2}\right].$$

(39)

From the deduced results, both the orthogonally polarized Doppler signals (Eqs. 38 and 39) give the same overall Doppler shift, and more importantly, its signs can be distinguished through the RPDs between these two orthogonally polarized Doppler signals.

If there were two specific helical phase masks ($-\ell_1$ and $-\ell_2$), one could get two Doppler shifts

$$\Delta\omega_j = \ell_j \Omega + 2kv - \frac{1}{2}(\alpha^2 + \beta^2)kv, \tag{40}$$

where $j=1$ and 2, and,

$$\text{sign}(\Delta\omega_j) = \begin{cases} +, & \Delta\varphi_j = \pi/2 \\ -, & \Delta\varphi_j = -\pi/2 \end{cases}. \tag{41}$$

Therefore, the multi-dimensional velocities ($v_z$ and $\Omega$) of the moving rough surface could be retrieved by solving these two sign-distinguishable Doppler shifts (Eqs. 40 and 41) as linear equations in two unknowns. Note that the Doppler signal that is converted to circular SoP by QWP after scattered from the moving rough surfaces in this case is equivalent to the case where the measuring light is converted to circular SoP in advance before illuminating the moving objects (see Case one in main text for detecting the moving particle). Both implementations can be used to produce RPDs between orthogonally polarized Doppler signals so as to distinguish the signs of Doppler shifts.

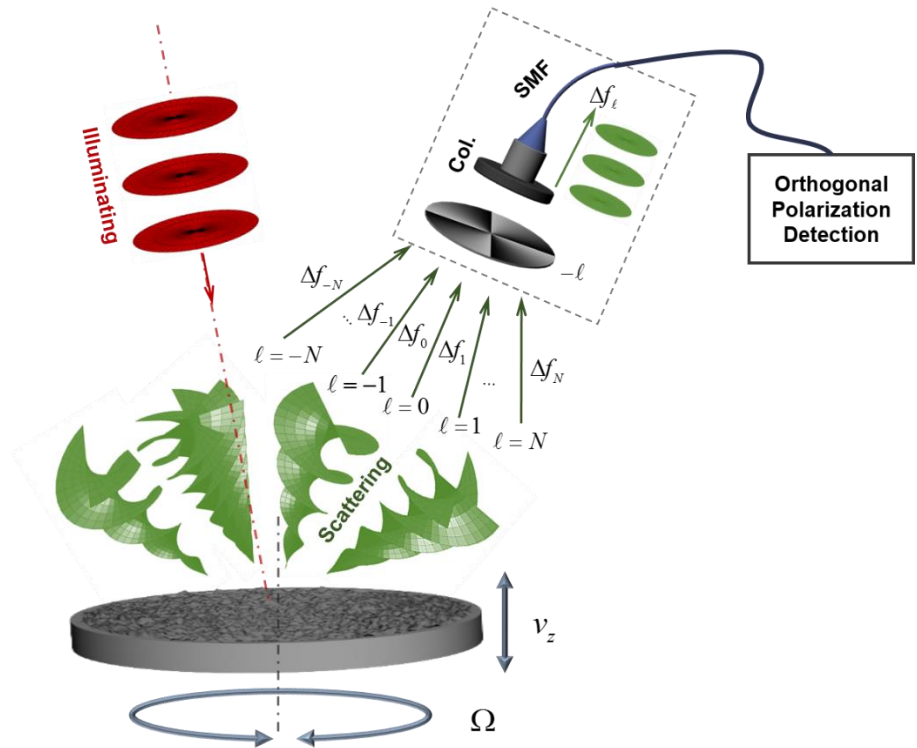

**Fig. S16 Conceptual diagram to extract the sign-distinguishable Doppler shift component to measure multi-dimensional movement of a moving rough surface using structured light** Col.: collimator; SMF: single-mode fiber. The combination of helical phase mask, collimator and SMF (shown as a box) is used for filtering desired twisted mode component to get the overall Doppler shift. The orthogonal polarization detection (similar to Case one in the main text) enables distinguish the signs of this Doppler shift.